\documentclass[%
 aip,
 amsmath,amssymb,
 reprint,%
]{revtex4-1}

\usepackage{graphicx}
\usepackage{dcolumn}
\usepackage{bm}
\usepackage{tabularx}
\usepackage[utf8]{inputenc}
\usepackage[T1]{fontenc}
\usepackage{mathptmx}
\usepackage{etoolbox}
\usepackage{booktabs}
\usepackage{siunitx}
\usepackage{float}
\usepackage{easyReview}
\usepackage{makecell}
\usepackage[normalem]{ulem}
\usepackage{longtable}
\makeatletter
\def\@email#1#2{%
 \endgroup
 \patchcmd{\titleblock@produce}
  {\frontmatter@RRAPformat}
  {\frontmatter@RRAPformat{\produce@RRAP{*#1\href{mailto:#2}{#2}}}\frontmatter@RRAPformat}
  {}{}
}%
\makeatother
\begin{document}

\preprint{AIP/123-QED}

\title[]{Regulation of droplet size and flow regime by geometrical confinement in a microfluidic flow-focusing device}
\author{Somasekhara Goud Sontti}
\author{Arnab Atta}%
 \email{arnab@che.iitkgp.ac.in}
\affiliation{Multiscale Computational Fluid Dynamics (mCFD) Laboratory, Department of Chemical Engineering, Indian Institute of Technology Kharagpur, Kharagpur, West Bengal 721302, India}


\begin{abstract}
\section*{ABSTRACT}

We have developed a coupled level set and volume of fluid (CLSVOF) based computational fluid dynamics (CFD) model to analyze the droplet formation mechanism in a square flow\textendash focusing microchannel. We demonstrate a flexible manipulation of droplet formation and flow regime based on the modified flow\textendash focusing microchannel with a constricted orifice. Furthermore, we have systematically studied the influence of geometrical confinement, flow rate, and interfacial tension on the droplet formation regime, length, volume, velocity, and shape. Three different flow regimes namely squeezing, dripping, and jetting are observed, and the flow regime maps are formulated based on the Reynolds and capillary numbers. After an extensive numerical investigation, we describe the boundaries between different regimes. Droplet shape is also quantified based on the deformation index value. Plug\textendash shaped droplets are observed in the squeezing regime, and near\textendash spherical droplets are found in the dripping and jetting regimes. Our study provides insights into the transition of a regime under various geometrical confinement and fluid properties. The results reveal that the modified flow\textendash focusing microchannel can substantially enhance dripping while decreasing the squeezing regime, which is of paramount importance from the standpoint of producing high throughput stable and monodisperse microdroplets. Eventually, this work emphasizes on the importance of geometrical confinement, fluid properties, and flow condition on the droplet formation process in a flow\textendash focusing microchannel that can effectively provide helpful guidelines on the design and operations of such droplet\textendash based microfluidic systems.
\end{abstract}

\maketitle

%
%

\section{Introduction}
The droplet microfluidics method has proven effective in chemical, biological, biomedical, and industrial processes over the past two decades.\cite{huebner2008microdroplets,zhu2017passive, yao2015review} Microfluidic droplet technology is attracted by its high surface\textendash to\textendash volume ratio and its excellent heat and mass transfer properties at a micro\textendash scale.\cite{theberge2010microdroplets, song2006reactions,abdollahi2017fluid} In turn, forming stable droplets is an essential component of droplet microfluidics. In addition to size, shape, and monodispersity, the droplets must meet the desired properties. Understanding the hydrodynamic parameters, e.g., droplet length, velocity, volume, and shape, is essential to understanding the microchannel performance.\cite{shang2017emerging, nguyen2013design} The uniform size of droplets is also required for various applications, including point\textendash of\textendash care diagnostics, drug delivery, cell/molecule compartmentalization, and diagnostic testing.\cite{guo2012droplet,teh2008droplet} Nevertheless, it is crucial to manipulate and generate stable droplets to ensure controlled and predictable results. Detailed knowledge of the droplet formation mechanism and its influencing parameters is necessary for a specific application.\cite{abiev2017hydrodynamics} For a formation of micro\textendash droplets, several microfluidic geometric configurations have been developed in microfluidic systems.\cite{chong2016active} Among those most frequently used configurations are co\textendash flow, T\textendash  junction, and flow\textendash focusing. \cite{chong2016active,zhu2017passive,kumari2022insights} The flow\textendash focusing microchannel is one of the most typically used microfluidic geometries for producing droplets, allowing high\textendash throughput production of monosized droplets by rupturing the dispersed phase symmetrically at the cross\textendash junction.\cite{anna2003formation,wu2013ferrofluid,tan2008drop,du2017self,sattari2017hydrodynamics}  Therefore, it is necessary to have a comprehensive understanding of monosized droplets for the effective control and utilization of droplet\textendash based flow. 

\citet{moon2016water} investigated water\textendash in\textendash water (\textit{W/W}) emulsions in flow\textendash focusing microchannel and studied the role of interfacial tension and viscosity on droplet formation. \citet{li2013} studied the synthesis of copper nanoparticles using the droplets generated within the flow\textendash focusing reactor. They reported that nanoparticle size was strongly controlled by altering the flow rate condition.~\citet{qi2021effect} explored the effect of nanoparticle surfactants on droplet formation in a flow\textendash focusing microchannel. Their results showed that droplet size was strongly controlled by the nanoparticle size. Moreover, their proposed theoretical model for predicting droplet size exhibited a good agreement with the experimental data. \citet{cheng2017janus} performed experimental investigations on Janus droplet formation and arrangements in the flow\textendash focusing microchannel. Their results indicate that flow rates significantly affect the size and frequency of Janus droplets. Recently, \citet{sundararajan2018engineering} experimentally investigated polymeric Janus particle formation for drug delivery applications. \citet{sattari2017hydrodynamics} experimentally analyzed the mass transfer characteristics in a liquid\textendash liquid system using three different aspect ratio microfluidic junctions. The chemical reaction between sodium hydroxide and acetic acid was performed in the microchannel. Their results showed a substantially higher mass transfer coefficient in flow\textendash focusing microchannel due to more significant movement of the biphasic interface and higher mixing within slugs. 

Few works were devoted to the formation mechanism by altering geometrical configurations at the cross junction. \citet{du2016breakup12} studied the formation of droplets in the flow\textendash focusing microchannel with symmetrical and asymmetrical rupture at the cross junction of the microchannel. It was evident from their study that the geometrical confinement strongly influenced the dispersed phase rupture phenomena. \citet{gulati2016microdroplet} analyzed droplet formation in a symmetric system by rounding four corners, and also in an asymmetric system by rounding two corners. Their results showed that in all the cases, monodisperse droplet was formed but with the largest droplets being produced at the junctions with the largest rounding. Droplet frequency, pinch\textendash off position (i.e., where dispersed phase thread breaks up completely and forms a droplet), and droplet size were well controlled by rounding the four corners of the flow\textendash focusing microchannel. Besides, the numerical results were also presented for single\textendash phase flow, where maximum velocity was witnessed in flow\textendash focusing junction with asymmetric rupture. \citet{amstad2017parallelization} delineated the influence of continuous phase shearing orientation in the flow\textendash focusing microchannel. \citet{li2021computational} numerically studied multi\textendash cell couple droplet generator system for high throughput production of microdroplets. They explored the coupling effects of flow parameters on the droplet characteristics and mechanisms in a multi\textendash cell parallel geometry. They reported that the droplet formation frequency increased with increasing continuous\textendash phase flow velocity while the size decreased.

Few numerical works reported on droplet formation in a flow\textendash focusing device using various computational approaches such as volume\textendash of\textendash fluid (VOF),\cite{chen2020modeling, roodan2020formation,ding2020droplet,khater2019dynamics} Level\textendash set (LS),\cite{mehraji2021flow,lan2014cfd}, Phase\textendash field, \cite{bai2017three} and Lattice Boltzmann method (LBM).\cite{gupta2016lattice} Coupled Level Set and Volume\textendash of\textendash Fluid (CLSVOF) method  was successfully implemented in various applications in droplet deformation, \cite{guan2019deformation} interface evolution of a droplet impact in the liquid film,\cite{bao2022interface} interfacial heat transfer and phase change applications,\cite{yang2022investigations} bubble formation,\cite{cao2020bubble} droplet coalescence,\cite{kwakkel2013extension,kagawa2014} bubble rise,\cite{ma2018numerical} and demulsification.\cite{mino2016numerical} In our previous work, the CLSVOF method was successfully implemented for low capillary numbers droplet and bubble formation in a microchannel.\cite{sontti2019numerical,sontti2018formation} This method resolves the nonphysical velocities near the interface, known as \textit{spurious currents}.\cite{popinet1999,harvie2006analysis} Therefore, in this study CLSVOF method is considered to investigate the droplet formation in a modified flow\textendash focusing microchannel.     

Nevertheless, the numerical understandings of droplet formation and its dynamics in a modified flow\textendash focusing microchannel with geometrical confinement are still insufficient. Moreover, all the aforementioned studies on droplet formation and flow regimes were reported in the standard flow\textendash focusing microchannel. Prior to designing microchannels for droplet flow regimes, it is necessary to have an understanding of hydrodynamic parameters such as droplet length, velocity, film thickness, and pressure drop. In addition, geometrical parameters are essential in understanding the droplet formation, shape, and its volume. To the best of our knowledge, droplet formation and flow regime maps based on geometric configurations are not yet explored. This study presents a comprehensive numerical analysis of droplet flow formation in a modified flow\textendash focusing microchannel with an orifice width and length. We have systematically investigated droplet formation at different continuous flow rates and interfacial tension. We have also developed flow regime maps using a modified flow\textendash focusing microchannel with geometric confinement. Our CLSVOF model elaborates the flow and pressure fields in the microchannel, which are not easily attainable by the experiments.  

This article contains the following structure. In Sec.~\ref{Problemformulation}, a brief description of the problem formulation and geometric configurations with dimensions is presented. In Sec. \ref{CLSVOF}, we discuss the interface capture method and governing equations. In Sec.\ref{Model}, we briefly describe the details of our numerical methodology settings and grid independence details. We then provide experimental validation of our numerical model. In Sec.\ref{Results}, we present and discuss the effects of the various constriction width, length, continuous phase flow rate, and interfacial tension on droplet formation and flow regimes at flow\textendash focusing geometry. Finally, we conclude our study with some concluding remarks in Sec.\ref{conclusions}.  

\section{Problem formulation} \label{Problemformulation}
\begin{figure*}[]
	\centering
	\includegraphics[width=0.80\textwidth]{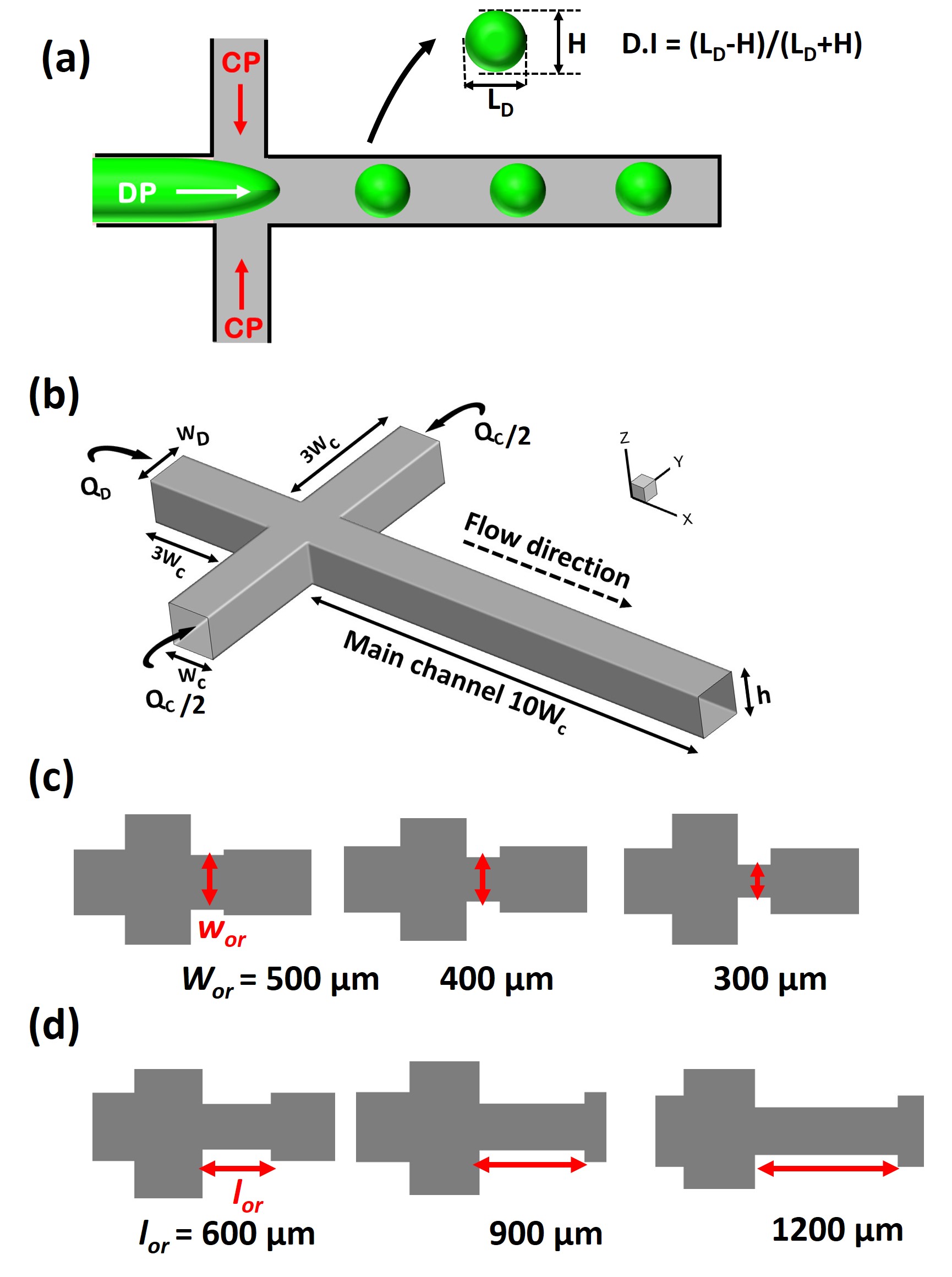}
	\caption{\label{fig:MD1} (a) Schematic of the standard flow\textendash focusing microchannel where CP: continuous phase flow rate and DP: dispersed phase flow rate, (b) three dimension view of computation domain with channel dimensions, (c) modified flow\textendash focusing microchannels with different orifice width for a fixed orifice length of 300 $\mu$m, and (d) modified flow\textendash focusing microchannels with different orifice length for a fixed orifice width of 300 $\mu$m}. 
\end{figure*}

Fig.~\ref{fig:MD1}a illustrates a planar view of the flow\textendash focusing microchannel. 
In the main channel, the dispersed phase (i.e., silicone oil) was fed, while the continuous phase (i.e., water) was pumped through the two side channels. In a flow\textendash focusing microchannel, both fluids meet at the cross junction. As a result of the interplay between different forces, oil\textendash in\textendash water (\textit{O/W}) droplets are formed at the downstream of the microchannel, as depicted in Fig.~\ref{fig:MD1}a.  A three\textendash dimensional (3D) standard configuration of the flow\textendash focusing microchannel with a square cross\textendash section having 600~$\mu$m $\times$ 600~$\mu$m height ($h$) is shown in Fig.~\ref{fig:MD1}b. The dimensions of continuous and dispersed phase inlets are 3$W_c$ ($W_c=W_d=W$) each, and the length of the main channel is 10$W_c$, as also shown in Fig.~\ref{fig:MD1}b. In this study, the standard flow\textendash focusing microchannel with simple modifications at the cross junction is considered. Various configurations of flow\textendash focusing microchannels by altering the constriction width and length are investigated, and the influence of constriction width and length are explored. Fig.~\ref{fig:MD1}c shows configurations with different constriction width ($w_{or}$) varying from 500 $\mu$m to 300 $\mu$m when the constriction length ($l_{or}$) was kept constant at 300 $\mu$m. Similarly, configurations with different constriction length are shown in Fig.~\ref{fig:MD1}d by altering the orifice length ($w_{or}$) from 600 $\mu$m to 1500 $\mu$m for the constant width ($w_{or}$) of 300 $\mu$m). A normalized orifice width/length is defined as the ratio between orifice width/length to channel depth (${w^{*}_{or}}= w_{or}/h$ and ${l^{*}_{or}}= l_{or}/h$).  As a matter of fact, our modified flow\textendash focusing microchannels and dimensional range can conveniently be fabricated for experiments (see supplement material of \citet{fallah2014enhanced} and \citet{baby2017fundamental}). The droplet length ($L_D$) is measured by drawing a horizontal reference line in the middle of the droplet, and length is considered as the distance from the rear to the nose of the droplet. In all cases, the average droplet length is considered from the available droplets in the microchannel after reaching a stable droplet formation regime. Droplet volume was calculated using the CFD post-processing tool. Its formation frequency ($f$) is also estimated based on the time required to form two successive droplets in a stable droplet formation regime. Droplet deformation index (DI) is calculated based on the length and width of the droplet, as shown in Fig.~\ref{fig:MD1}a. For $DI = 0$, the length and height of the droplet are identical, which indicates perfectly spherical droplet. However, $DI > 0$ indicates the deformed droplet with a length greater than its height.

\section{Coupled Level Set and VOF (CLSVOF) method}
\label{CLSVOF}
Over the past few decades, several interface-capturing models have been developed based on the Eulerian method. Among those, volume\textendash of\textendash fluid methods and the Level\textendash set method are extensively used for modeling interfacial flows \cite{hirt1981volume,sontti2017cfd,sontti2017cffd,sontti2019cfd,sussman1994level}. Nonetheless, the VOF method results in spurious currents despite being mass conservative.\cite{sonttiunderstanding} As a result of an unbalanced representation of surface tension forces and pressure variations across the interface, spurious currents are developed. Conversely, the LS method allows for smoother tracking. However, it has the drawback of mass conservation. \citep{suss-1994} In the LS method, a signed distance function is used to identify the phases. The Coupled Level Set and Volume\textendash of\textendash Fluid (CLSVOF) method was successfully implemented to capture the flow, evolution, and topology changes of an interface between two incompressible, immiscible viscous liquids subject to low capillary numbers (Ca) in our previous works.\cite{sontti2018formation,sontti2019numerical,sontti2019cfd} Therefore, in this study, a combined LS/VOF method (CLSVOF) is adopted to overcome the deficiencies of both LS and VOF methods for incompressible microfluidic flow with constant fluid properties. \cite{suss-1994,sontti2019numerical,sontti2018formation,sonttiunderstanding}     
\subsection{Fluid flow governing equations}

In the CLSVOF approach, the flow of immiscible fluids and laminar flow is described by a single set of conservation equations.

\textbf{Equation of continuity:} 

\begin{equation}
	\label{eq:mass_eqn}
	\nabla.   \vec{ U }  =0
\end{equation}

\textbf{Equation of motion:}

\begin{equation}
	\label{eq:mom_eqn}
	\rho \frac{ \partial ( \vec{ U })}{ \partial t} + \rho \nabla.( \vec{ U } \vec{ U }) = - \nabla P + \nabla. \eta (\nabla \vec {U} + \nabla { \vec {U} } ^{T}) + \rho g + \vec{ F}_{SF}
\end{equation}
where $\vec{U }$, $\rho $, $\eta $, and $P$ are denotes the velocity, density of fluid, dynamic viscosity of fluid, and pressure, respectively. The interfacial tension force is denoted by $\vec{ F}_{SF}$. 

\subsection{Interface capturing method} 
In the LS method, the fluid phase is described by the sign of the LS function ($\varphi$), which is a function of position ($\vec{\chi}$) and time ($t$). The LS function ($\varphi$) acts as the signed distance from the fluid-fluid interface, and is calculated as follows~\citep{suss-1994}:  
\begin{equation}
	\frac{\partial   \varphi }{\partial t}  +  	\vec{U} .  \nabla  \varphi=0  
\end{equation}
$\vec{U}$ and $\varphi$ are the velocity and LS functions, respectively. A level\textendash set function is a signed distance measure between positive and negative magnitudes of $\textit{d}$. The shortest distance from the interface to fluid denotes with $\textit{d}=d(\vec{r})$, at a point $\vec{r}$ and time $\textit{t}$.

\begin{equation}
\varphi(\vec{r}, t) =
\begin{cases}
d & \text{if } r \text{ is in the fully oil region},\\
0 & \text{if } r \text{ is in the oil–water interface},\\
-d & \text{if } r \text{ is in the fully water region}.
\end{cases}
\end{equation}

In this work, the fluid density and viscosity values are constant. In the fluid domain, two different values are assigned based on the sign of the LS function. These properties are subsequently estimated utilizing the smoothed Heaviside function  ($\textit{H}(\varphi)$) to determine continuous variation along with the interface.~\citep{sussman2000}. 

\begin{equation}
	\rho (\varphi )=\textit{H}(\varphi)\rho_{2}+(1-\textit{H}(\varphi))\rho_{1}  
\end{equation} 
\begin{equation}
	\eta(\varphi)=\textit{H}(\varphi)\eta_{2}+(1-\textit{H}(\varphi))\eta_{1}
\end{equation}

A smoothed Heaviside function ($\textit{H}(\varphi)$) can be expressed as follows:

\begin{equation}
H(\varphi) =
\begin{cases}
0, & \text{if } \varphi < -\epsilon,\\[6pt]
\dfrac{1}{2} \left[ 1 + \dfrac{\varphi}{\epsilon} + \dfrac{1}{\pi} \sin\!\left( \dfrac{\pi \varphi}{\epsilon} \right) \right], & \text{if } |\varphi| \leq \epsilon,\\[6pt]
1, & \text{if } \varphi > \epsilon.
\end{cases}
\end{equation}

where \textit{$\epsilon$} is the interface thickness.\\

\textbf{Surface tension modeling:}
The volumetric surface tension force ($\vec{F}_{SF} $ in Eq.\ref{eq:mom_eqn}) in CLSVOF is computed via a modification of the continuum surface force (CSF) model \citep{brack-1992} as follows~\citep{sussman1999}:

\begin{equation}
	\vec{ F}_{SF} = \sigma  \kappa ( \varphi ) \delta ( \varphi ) \nabla \varphi 
\end{equation}

where $\kappa ( \varphi )$ and $\delta ( \varphi )$  are the interface curvature and the smoothed Dirac delta function, respectively, defined as:

\begin{equation}
	\kappa ( \varphi )= \nabla .  \frac{ \nabla  \varphi }{ |  \nabla  \varphi  | } 
\end{equation}

\begin{equation}
\delta(\varphi) =
\begin{cases}
0, & \text{if } |\varphi| \geq \epsilon,\\[6pt]
\dfrac{1}{2\epsilon} \left( 1 + \cos\!\left( \dfrac{\pi \varphi}{\epsilon} \right) \right), & \text{if } |\varphi| < \epsilon.
\end{cases}
\end{equation}


\section{Model implementation and validation}
\label{Model}
\subsection{Solver settings and boundary condition}
The aforementioned time\textendash dependent partial differential equations are solved using a finite volume method. For solving the momentum equation with pressure\textendash velocity coupling, the pressure\textendash splitting operators (PISO) algorithm is applied. LS and momentum equations are discretized using the second\textendash order upwind scheme. Piecewise linear interface construction (PLIC) is used to solve the volume fraction. The governing equations are then solved with a  time step of $10^{-7}$ and a fixed Courant number (Co = 0.25). A constant velocity boundary condition is applied at the inlet for both the continuous (water) and dispersed (oil) phases. At the channel outlet, a pressure boundary condition is specified. No\textendash slip boundary condition is applied at the solid wall. Initially, it is assumed that the continuous phase occupies the entire main channel, and the disperse phase develops the flow profile at the cross\textendash junction where two\textendash phases interact. In this work, the continuous phase is assumed to completely wet the channel wall, and the droplet formation occurs at the junction without spreading the channel wall. Consequently, we specified a static contact angle ($\theta$= 120\textdegree) in our model. 

\subsection{Grid independence analysis and model validation}
\begin{figure} [!h]
	\centering
	\includegraphics[width=0.5\textwidth]{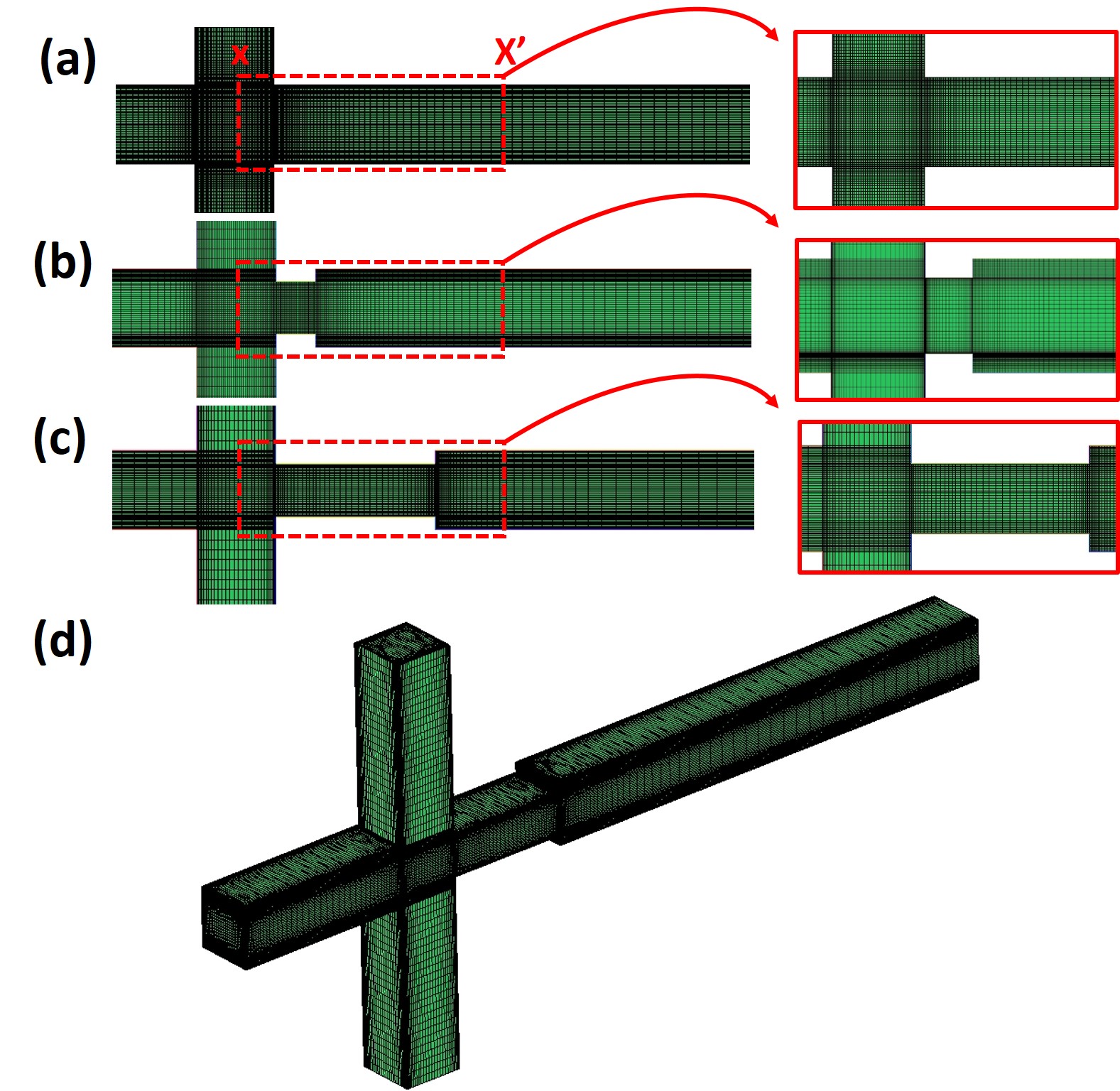}
	\caption{\label{fig:MDp1} Computational grid of different flow\textendash focusing  microchannels (a) standard flow\textendash focusing microchannel without orifice, (b) microchannel with smaller orifice width and length, (c) microchannel with largest orifice length, and (d) three dimensions view computational grid for the largest orifice length. Mesh refinement near the wall and orifice region  $X$\textendash $X^{'}$ is magnified in the insets.}
\end{figure}

At first, grid convergence analysis was conducted for the geometric configurations to assess the resolution of the interface, accuracy, and excellent results. The structured hexahedral grid is employed for meshing the computational domain, as shown in Fig.~\ref{fig:MDp1}.   

Various mesh element sizes (i.e., 20, 30, 40, and 50 $\mu$m) are compared to analyze grid sensitivity on droplet length. The optimum mesh element size of 30 $\mu$m  is considered in the core region for all the geometries with near-wall refinements. Along the constriction orifice region, fine mesh is adopted as depicted in   Fig.~\ref{fig:MDp1}b, c. To capture the smooth interface and flow phenomena near the wall, mesh refinement is applied to all edges with a mesh element size of 1 $\mu$m as shown in Fig.~\ref{fig:MD1}d. The reader can also refer to our previous work \citet{sontti2019numerical}, where the liquid film thickness is successfully captured in a 3D square microchannel (see Fig.~5 of \citet{sontti2019numerical}) with mesh refinement. A similar mesh refinement strategy is employed in the present study.

\begin{figure*} []
	\centering
	\includegraphics[width=\textwidth]{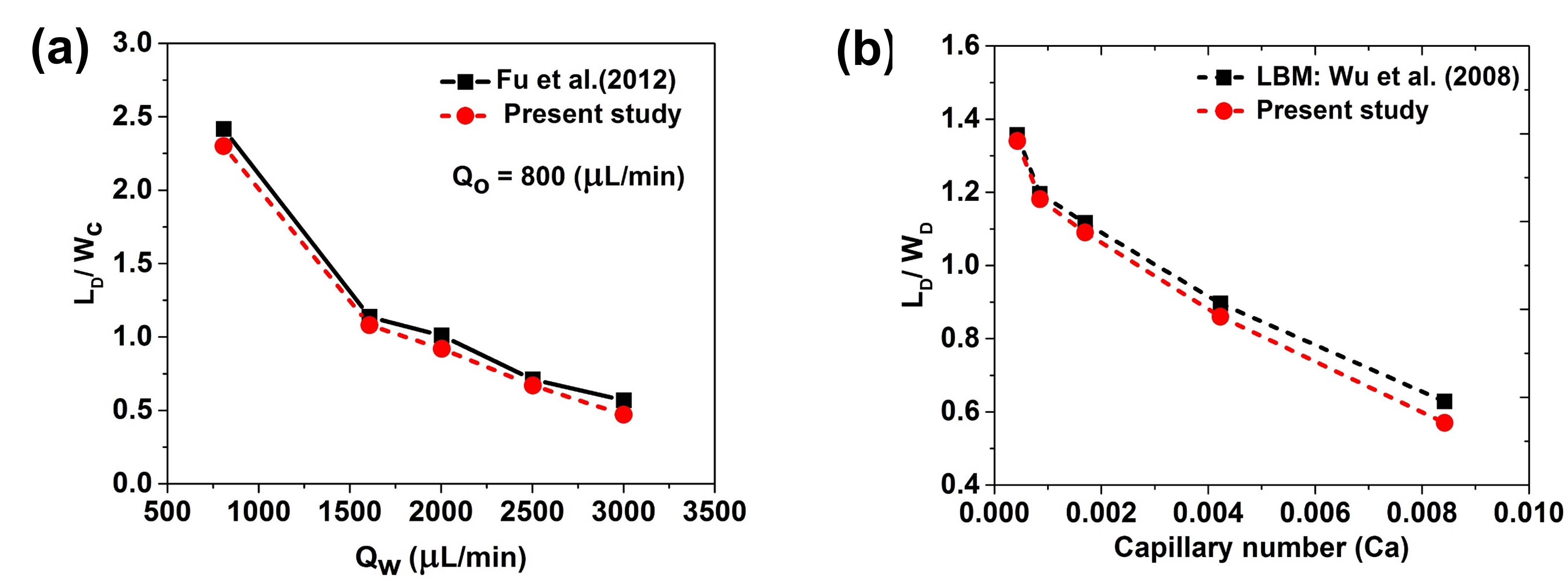}
	\caption{\label{fig:Mod1} Comparison of the dimensionless droplet length in a flow\textendash focusing microchannel against the (a) experimental results of \citet{fu-2012s} at fixed operating condition of $Q_o$= 400~$\mu$L/min, $\mu_o$= 11 mPa s, $\mu_o /\mu_w$ = 12, and $\gamma$ = 11.8 mN/m, and (b) LBM numerical results of \citet{wu2008three} at a fixed operating  condition of $U_o$ = 0.00084 m/s, $U_o/U_w$ = 0.33, $\mu_o$ = 24 mPa s, $\mu_o /\mu_w$ = 2.4.}
\end{figure*}

At first, we systematically investigated the efficacy of our developed CLSVOF CFD model by studying oil\textendash water flow in a flow\textendash focusing microchannel. Experimental results of \citet{fu-2012s} are compared with CFD results at the same operating condition and fluid properties. As shown in Fig.~\ref{fig:Mod1}a good agreement between CFD  and experiments is obtained in the dimensionless droplet determination as to the deviation of 4 \%. 
It is found that dimensionless droplet length decreases with an increase in continuous phase flow rate. The CFD model further verified the different water\textendash oil systems of \citet{wu2008three}. Fig.~\ref{fig:Mod1}b shows the quantitative comparison model prediction for a dimensionless droplet with LBM results reported by \citet{wu2008three} with a maximum deviation of 7\%. From this validated model, systematic numerical investigations are carried out to understand the role of confinement in a modified flow\textendash focusing microchannels and the impact of continuous phase flow rate and interfacial tension.         
\section{Results and discussion}
\label{Results}
\subsection{Effect of constriction width} 
\begin{figure*}[]
	\centering
	\includegraphics[width=\textwidth]{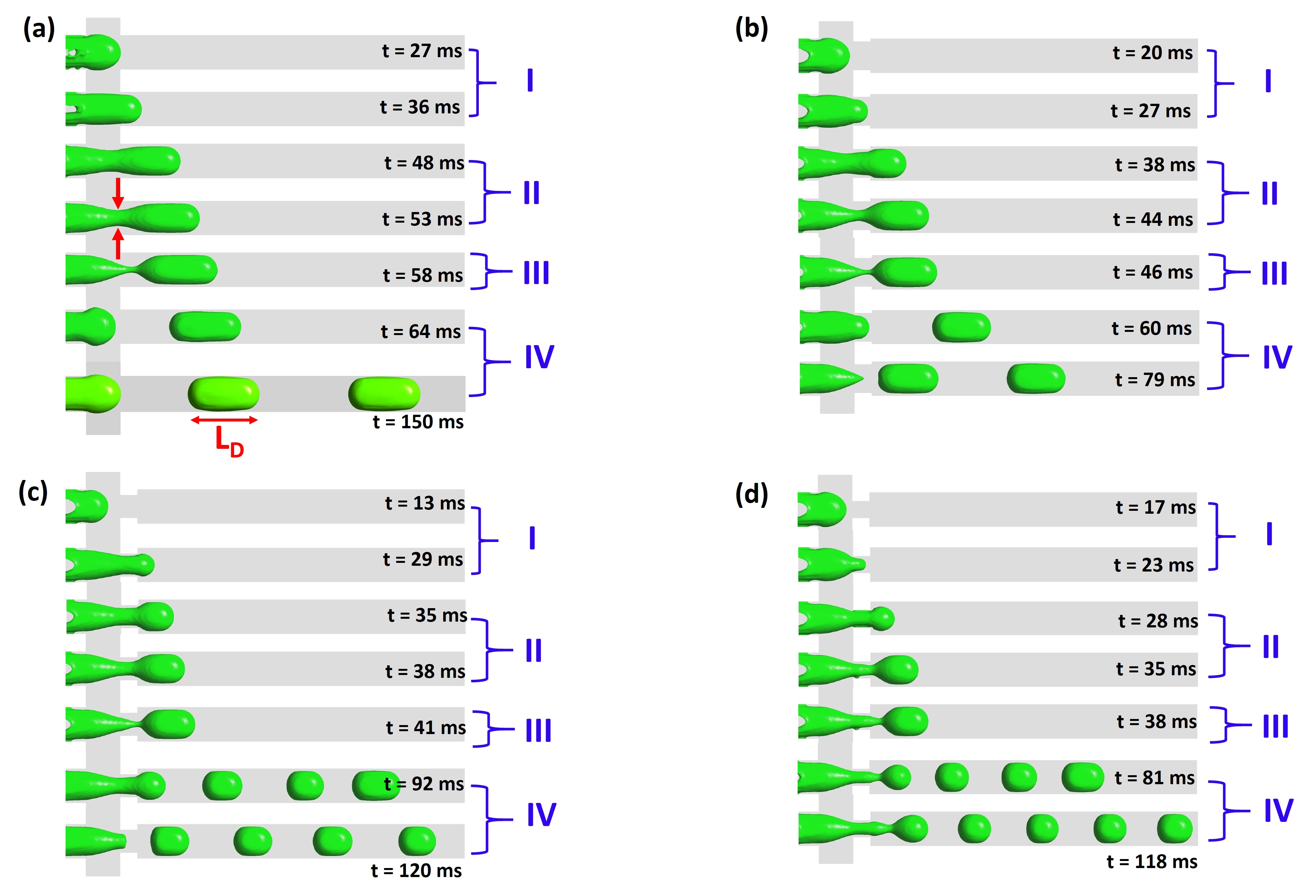}
	\caption{\label{fig:Cay1} Temporal evolution of droplet formation process in different stages: I\textendash expansion stage , II\textendash necking, III\textendash pinch\textendash off stage, and IV\textendash stable droplet formation regime. Droplet formation process in (a) standard flow\textendash focusing microchannel without orifice ${w^{*}_{or}}=1$, (b) modified flow\textendash focusing microchannel ${w^{*}_{or}}= 0.83$, (c) modified flow\textendash focusing microchannel ${w^{*}_{or}}= 0.66$, and (d) modified flow\textendash focusing microchannel ${w^{*}_{or}}= 0.5$ at a fixed operating condition of orifice length $l_{or}= 300~\mu m$, oil viscosity $\mu_o$ = 0.53 mPa s, $\mu_o /\mu_w$ = 0.59, interfacial tension $\gamma$ = 5.37 mN/m, $\theta$= 120\textdegree, $Q_w/Q_o=2$ and $Q_o$= 400~$\mu$L/min. }
\end{figure*}

Geometrical constriction width plays a crucial role in the droplet formation process. The study of constriction width on droplet formation in a modified flow\textendash focusing microchannel is useful for understanding how the channel constriction influences the formation process. In order to study the effect of constriction width on droplet formation, other flow rates and fluid properties were kept constant. For a deeper understanding of droplet formation, the temporal evolution of the droplet break process was analyzed as shown in Fig.~\ref{fig:Cay1}. 
The flow time $t$ is considered when the dispersed phase enters the cross\textendash junction. For all the cases, the droplet formation process in the squeezing and dripping regimes can be divided into four stages: I\textendash expansion stage, II\textendash necking, III\textendash pinch\textendash off, and IV\textendash stable droplet formation regime as depicted in Fig.~\ref{fig:Cay1}. In the case of a standard flow\textendash focusing microchannel without orifice (i.e., ${w^{*}_{or}}=1$), the squeezing regime was found to occur as illustrated in Fig.~\ref{fig:Cay1}a. 

At a fixed operating condition, shear force on the dispersed phase and the squeezing action of the continuous phase is considerably smaller compared to viscous and interfacial tension for low $Re$ flows. Under such scenario, the viscous and interfacial tension forces are responsible for the development of a longer thread and change of pinch-off position in a modified flow–focusing microchannel. In the cross\textendash junction, the forepart of the dispersed phase thread keeps expanding and remains spherical due to interfacial tension. As the dispersed phase volume increases in cross\textendash junction, the gap between the dispersed phase and corners of the microchannel narrows. In the expansion stage, the dispersed phase restricts the continuous phase flow downstream of the main channel as shown in Fig.~\ref{fig:Cay1}a at $t = 36$ms. This phenomenon is also similar to modified flow\textendash focusing microchannel cases. However, with the increase in the channel constriction width, the dispersed phase grows rapidly at cross\textendash junction, as shown in Fig.~\ref{fig:Cay1}d at $t$ = 23 ms. A significant change in flow phenomena at the cross\textendash junction and downstream of the channel is responsible for this.

As shown in Fig.~\ref{fig:Cay1}a at $t$ = 48\textendash 53  ms, the dispersed phase starts squeezing due to increased hydrostatic pressure inside the cross\textendash junction. Nevertheless, the observation of the necking stage is similar for all cases. Still, the temporal evolution time was decreased with an increase in constriction width as shown in Fig.~\ref{fig:Cay1}d at $t = 35$ ms. Once the hydrostatic pressure develops in the cross\textendash junction and  side channels are sufficient to squeeze, the dispersed phase pinch\textendash stage begins. The developed pressure is adequately considerable to dominate the viscous and interfacial tension in the pinch\textendash off stage. During the pinch\textendash stage, the dispersed phase thread tip gradually thins and blocks the cross\textendash junction of the main channel as shown in Fig.~\ref{fig:Cay1}a at $t = 53$ms. After reaching the convex interface of the dispersed thread is completely separated from the channel walls (Fig.~\ref{fig:Cay1}a at $t = 53$ms), the dispersed phase evolution enters the pinch\textendash off stage as depicted in Fig.~\ref{fig:Cay1}a at $t = 58$ms.  In a pinch\textendash off the stage, the thread is stretched downstream of a channel, and the dispersed phase neck continuously shrinks. After reaching the minimum neck size, the droplet formation process enters the pinch\textendash off stage. Note that, at the early stage of pinch\textendash off, the neck is out of contact with the solid walls and is surrounded by the continuous phase (Fig.~\ref{fig:Cay1}c). With increasing the channel constriction width, the dispersed phase rapidly reaches the pinch\textendash off stage as shown in Fig.~\ref{fig:Cay1}d at $t = 38$ ms. The flow phenomena then shift to a regime of stable droplet formation where continuous plugs and nearly spherical droplets are generated.  Our findings on formation mechanism and temporal evolution results show a reasonable agreement with literature data in a standard flow\textendash focusing \cite{wu2017role,yu2022experiment} and modified flow\textendash focusing systems \cite{du2017self,gupta2014droplet}.              

\begin{figure*}[]
	\centering
	\includegraphics[width=0.70\textwidth]{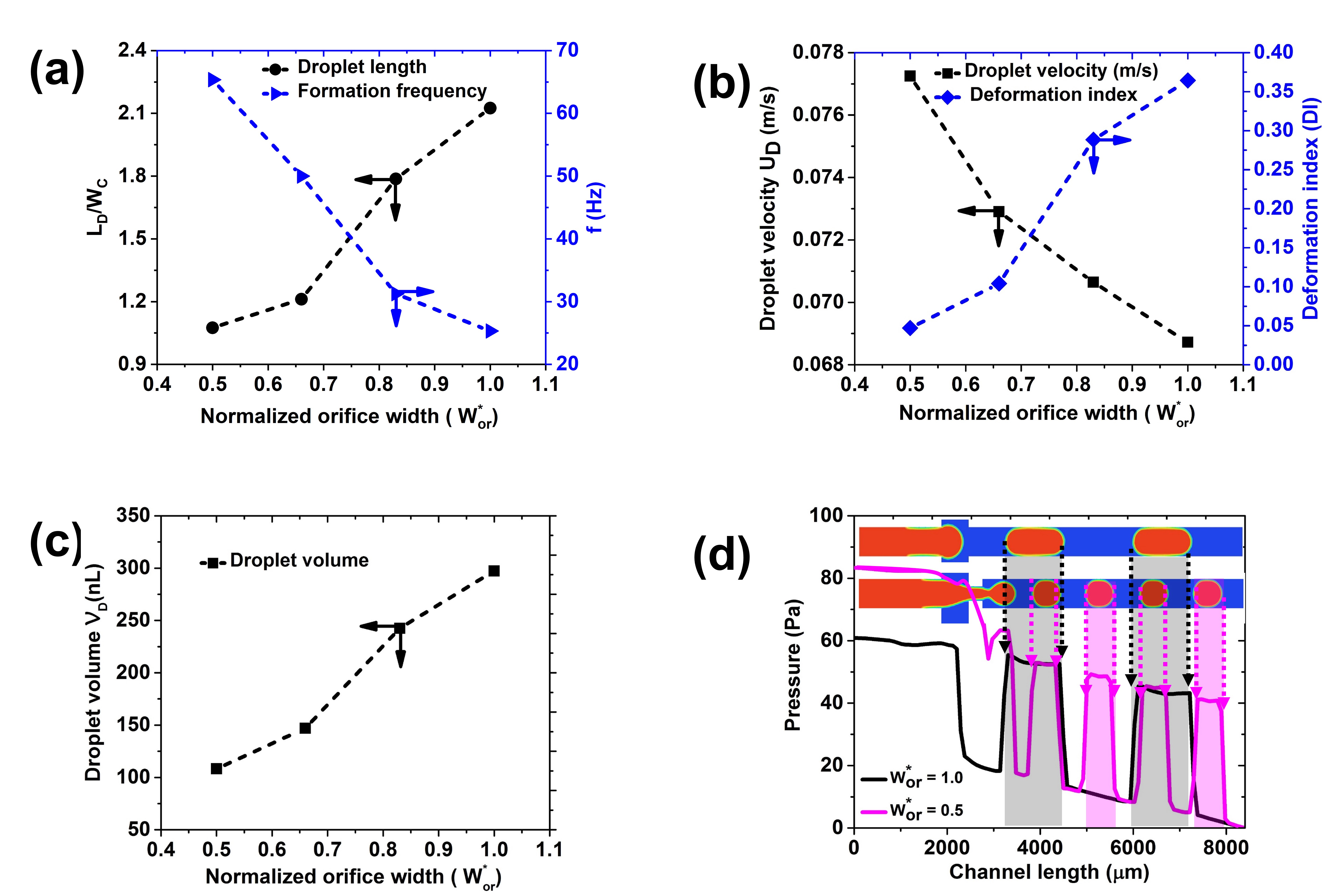}
	\caption{\label{fig:Orf1} Effect of orifice width on (a)  non\textendash dimensional droplet length and droplet formation frequency, (b) droplet velocity and droplet deformation index, (c) droplet volume, and (d) pressure profiles in the middle of the microchannel along the channel length at fixed operating condition of $\theta$= 120\textdegree, orifice length $l_{or}= 300~\mu m$,  oil viscosity $\mu_o$ = 0.53 mPa s, $\mu_o /\mu_w$ = 0.59, interfacial tension $\gamma$ = 5.37 mN/m, $Q_w/Q_o=2$ and $Q_o$= 400~$\mu$L/min. }
\end{figure*}

Fig.~\ref{fig:Orf1}a demonstrate the effect of the constriction width on non\textendash dimensional droplet length and droplet formation frequency at a fixed operating condition and flow rate. 
It is evident from Fig.~\ref{fig:Orf1}a that with increasing the constriction width, the non\textendash dimensional droplet length increases and formation frequency decreases. This is mainly due to an increase in shear force on the dispersed phase at the cross\textendash junction. In other words, the viscous force resulting from the continuous phase and the sudden change in geometrical width due to confinement facilitate the faster droplet formation. \cite{lashkaripour2019performance} In the case of normalized orifice width ${w^{*}_{or}}= 0.5$, the dispersed phase completely blocks the cross\textendash junction, and the dispersed phase expands downstream of the main channel. The developed pressure gradient leads to rapid smaller plug and near\textendash spherical droplet formation when orifice width ${w^{*}_{or}}< 1.0$. This is mainly ascribed to the change in confinement width, which alters the shear force on the dispersed phase. Interestingly, with increasing the orifice width ${w^{*}_{or}}$ from 0.5 to 1, the droplet size and frequency are significantly increased.

Fig.~\ref{fig:Orf1}b, shows the effect of constriction width on droplet velocity and droplet deformation index. It is found that with increasing the constriction width, droplet velocity decreases, and the deformation index increases. It is well known that the droplet velocity depends on the liquid film thickness and shape of the droplet. \cite{haase2017characterisation,sontti2019numerical} It can also be noted from Fig.~\ref{fig:Cay1} a\textendash d, the droplet shape and flow regime changes by altering constriction width. As a result, the droplet velocity decreases with increased constriction width. In the case of normalized orifice width ${w^{*}_{or}}= 0.5$, the flow regime is shifted to dripping from squeezing and also near\textendash spherical droplets are formed. Interestingly, when the orifice width ${w^{*}_{or}}= 0.5$, generated droplets deformation index is also smaller compared to other higher orifice width microchannels. Fig.~\ref{fig:Orf1}c demonstrates the effect of constriction width on droplet volume. With increasing the constriction width, droplet volume significantly increased due to a change in flow regime. Furthermore, the pressure variation for two different geometries analyzed along the center line of the microchannel is presented in Fig.~\ref{fig:Orf1}d. The pressure profile is a nearly identical trend for both cases. However, the pressure field in the dispersed phase is found to have a significant difference due to changes in flow regime and droplet shape. The  pressure at the inlet is different for ${w^{*}_{or}}= 0.5$ and ${w^{*}_{or}}= 0.1$. It is worth noting that the pressure profile of each peak indicates the droplet position and shape. Fig.~\ref{fig:Orf1}d shows the droplet position in the channel and corresponding  pressure profile for ${w^{*}_{or}}= 0.1$. 

\begin{figure*}[]
	\centering
	\includegraphics[width=0.5\textwidth]{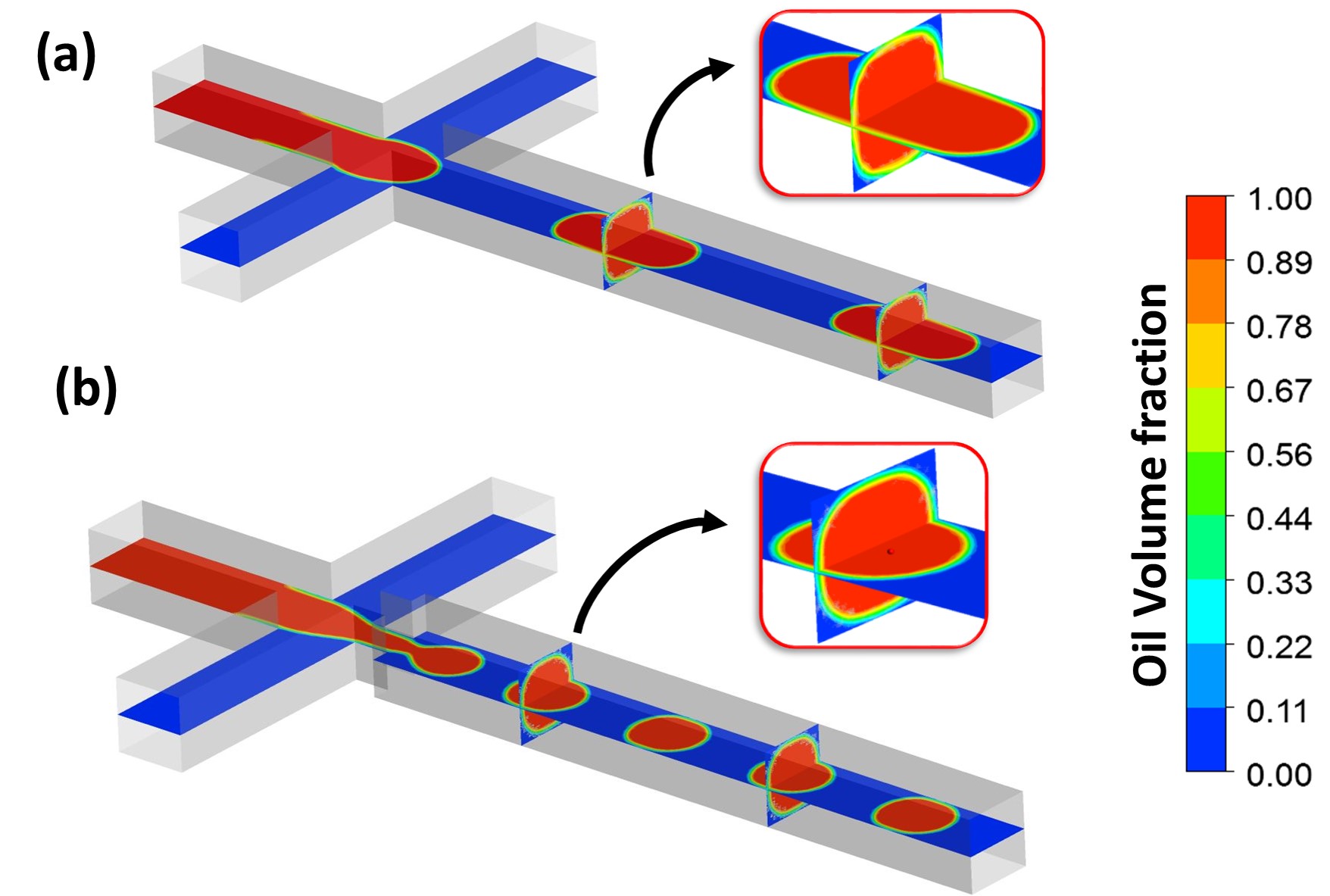}
	\caption{\label{fig:Film} Numerical visualization of volume fractions contours in middle of the microchannel for (a) standard flow\textendash focusing microchannel without orifice ${w^{*}_{or}}=1$ and (b) modified flow\textendash focusing microchannel ${w^{*}_{or}}= 0.5$ at a fixed operating condition of $\theta$= 120\textdegree, orifice length $l_{or}= 300~\mu m$,  oil viscosity $\mu_o$ = 0.53 mPa s, $\mu_o /\mu_w$ = 0.59, interfacial tension $\gamma$ = 5.37 mN/m, $Q_w/Q_o=2$ and $Q_o$= 400~$\mu$L/min (blue, water; red, oil). }
\end{figure*}

Fig.~\ref{fig:Film} shows the three\textendash dimensional visualization and volume fraction contour in the middle of microchannel for with orifice ${w^{*}_{or}}=1$  and   ${w^{*}_{or}}=0.5$. 
The flow regime changes from squeezing to dripping, and liquid film thickness is observed around the droplet. Magnified views are also inserted in Fig.~\ref{fig:Film} to visualize the liquid film thickness. From the magnified views, the dispersed phase is completely isolated from the solids wall, and the liquid is observed between the droplet and solid wall. In a square microchannel, the liquid film thickness differs from the channel's middle to the corners. This observation is similar to our previous study on three\textendash dimensional iso\textendash surface visualizations (see Fig.~5 of \citet{sontti2019numerical}). These results further establish the accuracy of our CLSVOF method in capturing the film thickness around the droplet.

\begin{figure}[!h]
	\centering
	\includegraphics[width=0.5\textwidth]{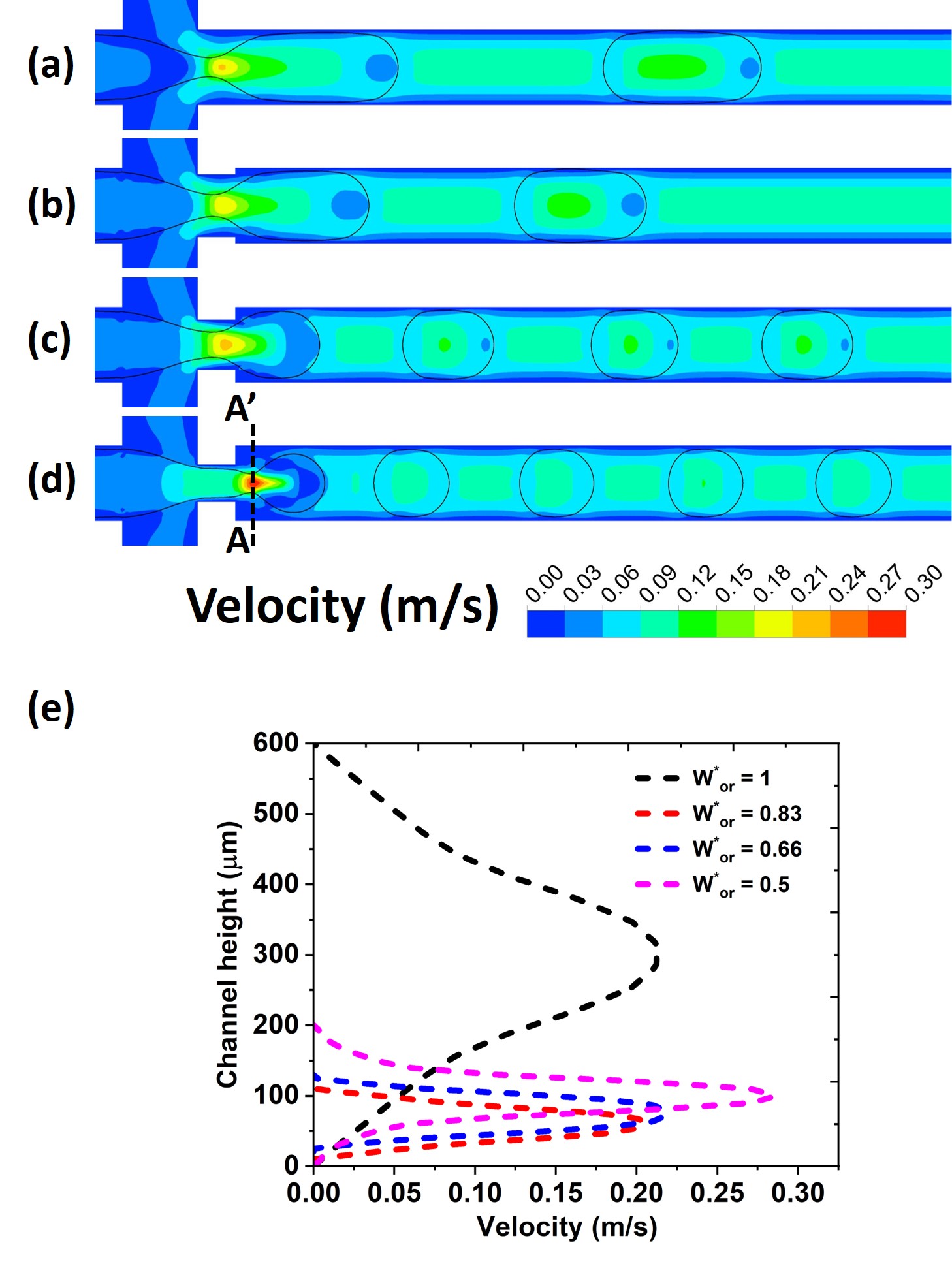}
	\caption{\label{fig:Va1} Effect of orifice width on velocity field variation in the microchannel for (a) standard flow\textendash focusing microchannel without orifice ${w^{*}_{or}}=1$ and modified flow\textendash focusing microchannel (b) ${w^{*}_{or}}= 0.83$, (c) ${w^{*}_{or}}= 0.66$, and (d) ${w^{*}_{or}}= 0.5$. (e) Velocity profiles at the pinch of position along reference line $A$ \textendash $A^{'}$ at a fixed operating condition of $\theta$= 120\textdegree, orifice length $l_{or}= 300~\mu m$,  oil viscosity $\mu_o$ = 0.53 mPa s, $\mu_o /\mu_w$ = 0.59, interfacial tension $\gamma$ = 5.37 mN/m, $Q_w/Q_o=2$ and $Q_o$= 400~$\mu$L/min.}
\end{figure}

Furthermore, the effect of the constriction width on the velocity field inside the microchannel is visualized as shown in Fig.~\ref{fig:Va1}. 
It is found that with increasing the constriction width, the velocity field on the dispersed phase neck significantly increases due to the constriction as shown in Fig.~\ref{fig:Va1}e. This is mainly ascribed to blocking the dispersed phase in the constriction and increasing the shear force on the dispersed phase. The dispersed is identified with a black color iso\textendash surface line from the velocity field visualizations. From Fig.~\ref{fig:Va1}, it is clear that the middle of the droplet velocity field is higher, which is closely similar to the liquid plug.
  
\subsection{Effect of constriction length}
\begin{figure*}[]
	\includegraphics[width=0.80\textwidth]{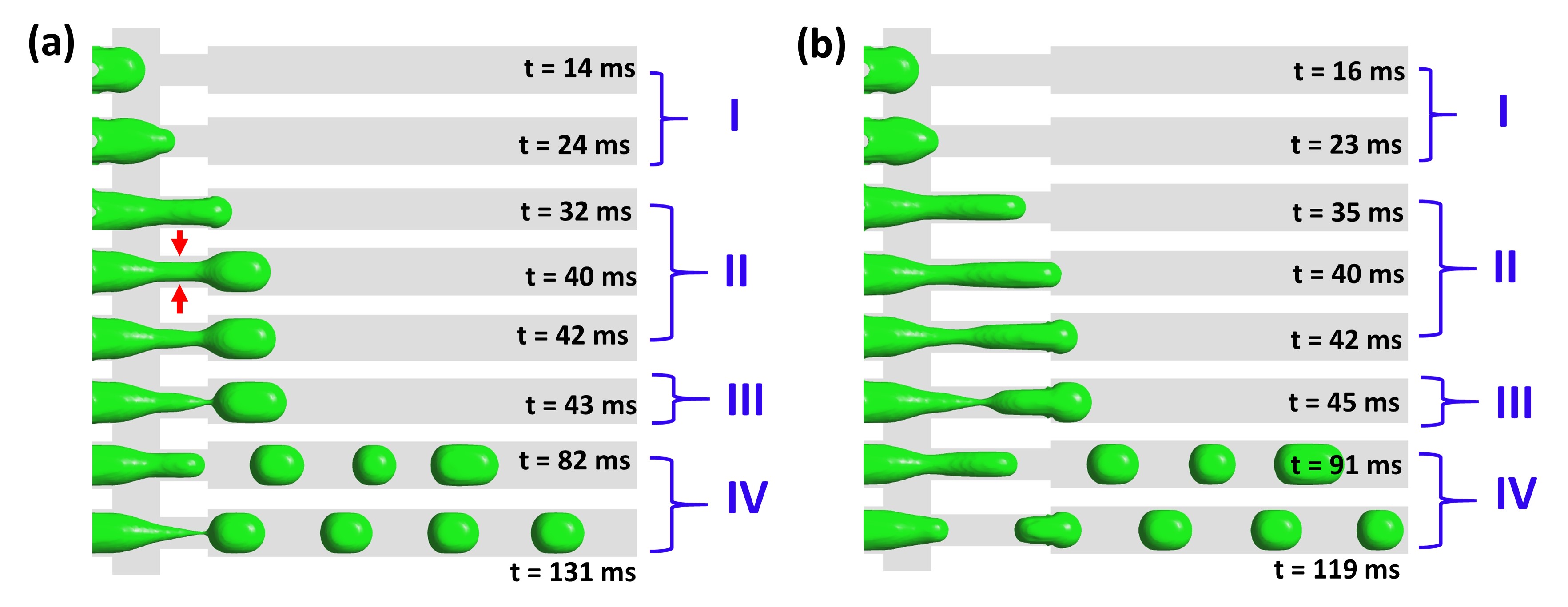}
	\caption{\label{fig:Ca1} Temporal evolution of droplet formation process in different stages: I\textendash expansion stage , II\textendash necking, III\textendash pinch\textendash off stage, and IV\textendash stable droplet formation stage. Droplet formation process in two different flow\textendash focusing microchannels with (a) normalized orifice length ${l^{*}_{or}}=1$, (b) normalized orifice length  ${l^{*}_{or}}= 2.5$ at a fixed operating condition of $\theta$= 120\textdegree, orifice width $w_{or}= 300~\mu m$,  oil viscosity $\mu_o$ = 0.53 mPa s, $\mu_o /\mu_w$ = 0.59, interfacial tension $\gamma$ = 5.37 mN/m, $Q_w/Q_o=2$ and $Q_o$= 400~$\mu$L/min. }
\end{figure*}

This section studies the effect of constriction length on droplet formation at a fixed operating condition and orifice width. Fig.~\ref{fig:Ca1} shows the temporal evolution of the droplet formation process for two different microchannels. Similar to our previous discussion, the droplet formation process can be divided into four stages. From the morphology evolution of the dispersed phase, it can be observed that the squeezing force on the dispersed phase is less with increasing the constriction length. As a result, the dispersed thread expands radially at the cross\textendash junction as shown in Fig.~\ref{fig:Ca1} and at $t=24$ ms.

  During the necking stage, the dispersed thread expands axially inside the constriction, and the necking of the dispersed thread ruptures symmetrically in the middle. Fig.~\ref{fig:Ca1}a demonstrates that the squeezing force on the dispersed phase prevails over the interfacial tension force at higher constriction lengths. However, the necking stage evolution is the same in both cases. The droplet formation process reaches the pinch\textendash off stage when the dispersed thread becomes thinner. As shown in Fig.~\ref{fig:Ca1}a, smaller orifice length of microchannel quickly reaches the pinch\textendash off stage. As depicted in Fig.~\ref{fig:Ca1}a and b, a stable droplet is formed when the dispersed phase pinches off in the middle of constricting. The dripping regime is noticeable in the range of constriction lengths and operating condition. In the dripping regime, stable droplet formation is observed and excellent monodispersity (see Fig.~\ref{fig:Ca1} at $t=131$ms).
  
  \begin{figure*} []
  	\centering
  	\includegraphics[width=\textwidth]{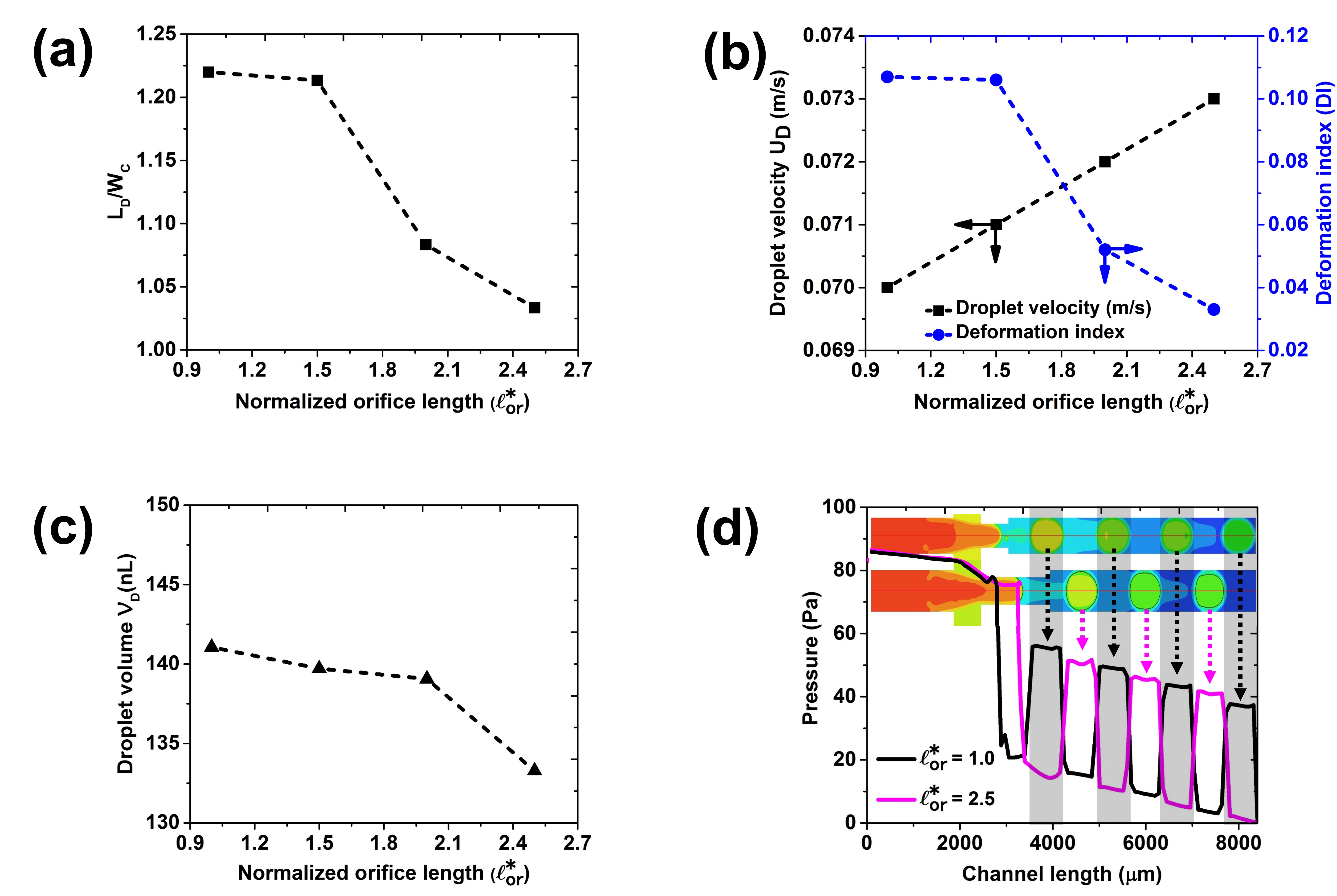}
  	\caption{\label{fig:Orfl1} Effect of orifice length on (a) non\textendash dimensional droplet length, (b) droplet velocity and droplet deformation index, (c) droplet volume, and (d) pressure profiles in the middle of the microchannel along the channel length at a fixed operating condition of $\theta$= 120\textdegree, orifice width $w_{or}= 300~\mu m$,  oil viscosity $\mu_o$ = 0.53 mPa s, $\mu_o /\mu_w$ = 0.59, interfacial tension $\gamma$ = 5.37 mN/m, $Q_w/Q_o=2$ and $Q_o$= 400~$\mu$L/min. }
  \end{figure*}
 
The influence of constriction length on droplet formation is quantified as shown in Fig.~\ref{fig:Orfl1}. We have found that non\textendash dimensional droplet length decreases with lengthening the constriction.
As demonstrated from Fig.~\ref{fig:Orfl1}b, droplet velocity increased linearly as the constriction length increased. This is primarily due to an increase in liquid film thickness and a change in the velocity field in the main channel. Also, a change in the velocity field after the constriction leads to accelerating of droplet velocity in the main channel.  Droplet shape is also quantified by the droplet deformation index, as shown in Fig.~\ref{fig:Orfl1}b. As the constriction length increases, a plug shape droplet transitions to a near\textendash spherical droplet. Droplet volume is also found decreased with constriction length, as shown in Fig.\ref{fig:Orfl1}c. Therefore, our findings revealed that droplet size and volume could be controlled with constriction length. Furthermore, the pressure profile middle of the microchannel is analyzed for two different orifice\textendash length microchannels. From Fig.~\ref{fig:Orfl1}d, it can be observed that the change in pressure along the microchannel length is negligible, but each peak magnitude is different due to the droplet location and size. For the normalized orifice length ${l^{*}_{or}}=1$ case, corresponding pressure contour is also shown as an inset in Fig.~\ref{fig:Orfl1}d. The dispersed phase partially blocks the main channel as it enters the cross junction. While it occupies the cross junction, the upstream and downstream pressure of the continuous phase increases.\cite{sontti2019numerical} Therefore, the pressure variation can be different in the cross\textendash junction depending on the orifice width and length configuration. Consequently, the droplet formation is driven by the pressure variation at the cross-junction between both continuous and dispersed phases.    

\begin{figure*} []
	\centering
	\includegraphics[width=0.8\textwidth]{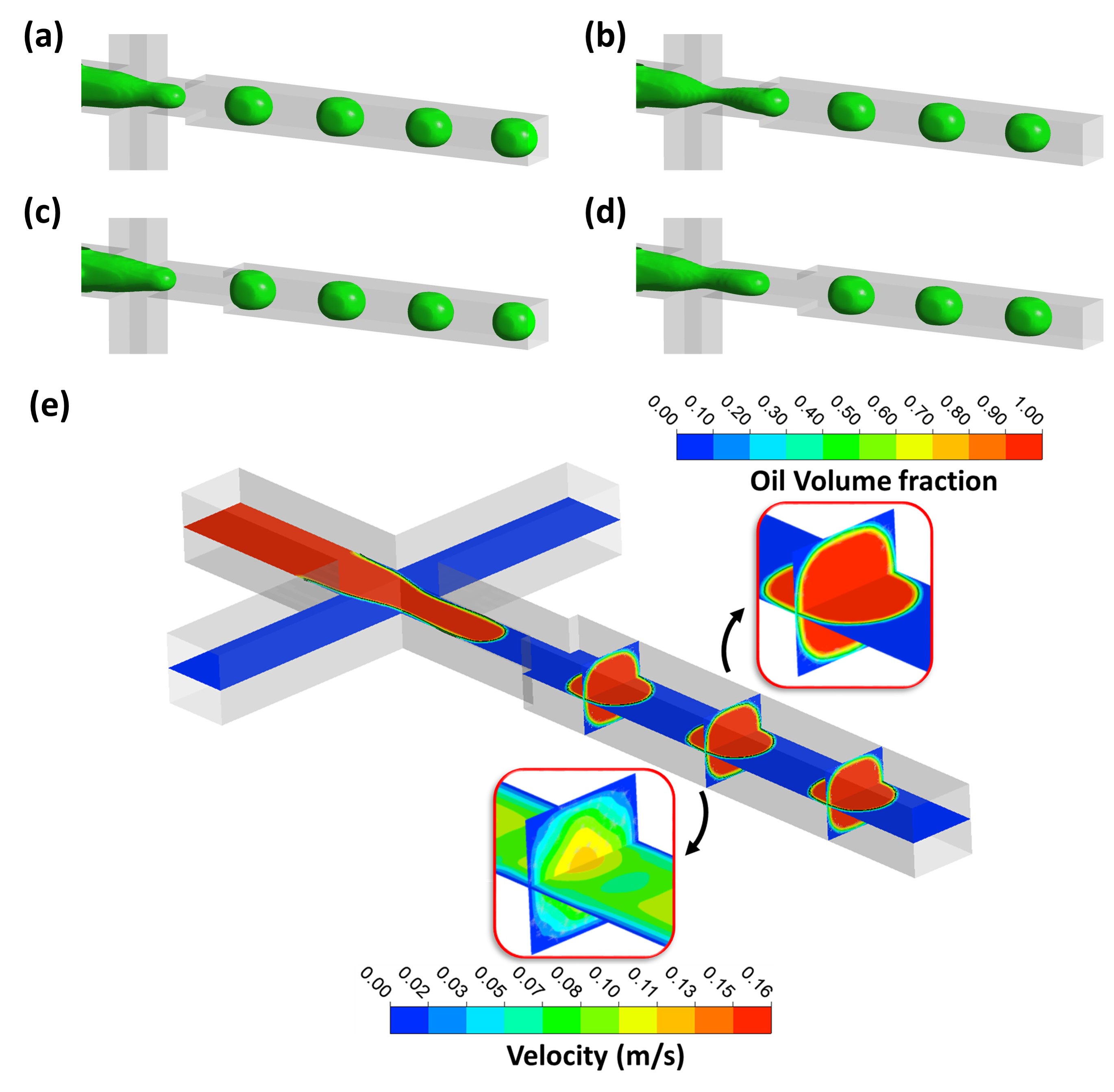}
	\caption{\label{fig:Sa1} Three\textendash dimensional iso\textendash surface view for different normalized orifice lengths (a) ${l^{*}_{or}}=1$, (b) ${l^{*}_{or}}=1.5$, (c) ${l^{*}_{or}}=2$, (d) ${l^{*}_{or}}=2.5$, and (e) volume fraction contour visualization in the middle of the microchannel: liquid fill thickness around the droplet and velocity field inside the droplet is amplified in the inset at a fixed operating condition of  $\theta$= 120\textdegree, $w_{or}= 300~\mu m$, normalized orifice length  ${l^{*}_{or}}= 2.5$,  oil viscosity $\mu_o$ = 0.53 mPa s, $\mu_o /\mu_w$ = 0.59, interfacial tension $\gamma$ = 5.37 mN/m, $Q_w/Q_o=2$, and $Q_o$= 400~$\mu$L/min (blue, water; red, oil).  }
\end{figure*}

Fig.~\ref{fig:Sa1}a\textendash d shows the three\textendash dimensional iso\textendash surface views of the dripping regime for different normalized orifice lengths for a fixed orifice width and operating condition. 
A clear distinction is observed with increasing the orifice length on droplet length and slightly changing to small plugs. These 3D flow visualization results reveal that the normalized orifice length did not strongly affect the flow transition for the range of operating condition. This is attributed to lower shear force on the dispersed phase, specifically inside constriction. Thus, viscous or interfacial tension has a higher impact on the dispersed phase, leading to stable droplets forming. Fig.~\ref{fig:Sa1}e demonstrates the liquid film thickness is visualization around the droplet for a higher orifice length. As shown in the inset of Fig.~\ref{fig:Sa1}e, vertical slices are created in the middle of the droplet. It is observed that thin liquid film is observed along the walls and film thickness enlarged in the four corners. This can be further explained by velocity field observation around the droplet as shown in the inset of  Fig.~\ref{fig:Sa1}e.
          
\subsection{Effect of continuous phase flow rate}
This section investigates the effect of continuous phase flow rate on droplet formation for four different microchannels with varying the orifice width while other fluid properties are constant. The dispersed phase flow rate and orifice length are constant for all cases. Continuous phase Reynolds numbers $\big(Re =  \frac{ W_{c}  U_{c}   \rho _{c}  }{  \mu_{c}  } \big)$ are used to quantify the results. 
\begin{figure*}[]
	\centering
	\includegraphics[width=0.8\textwidth]{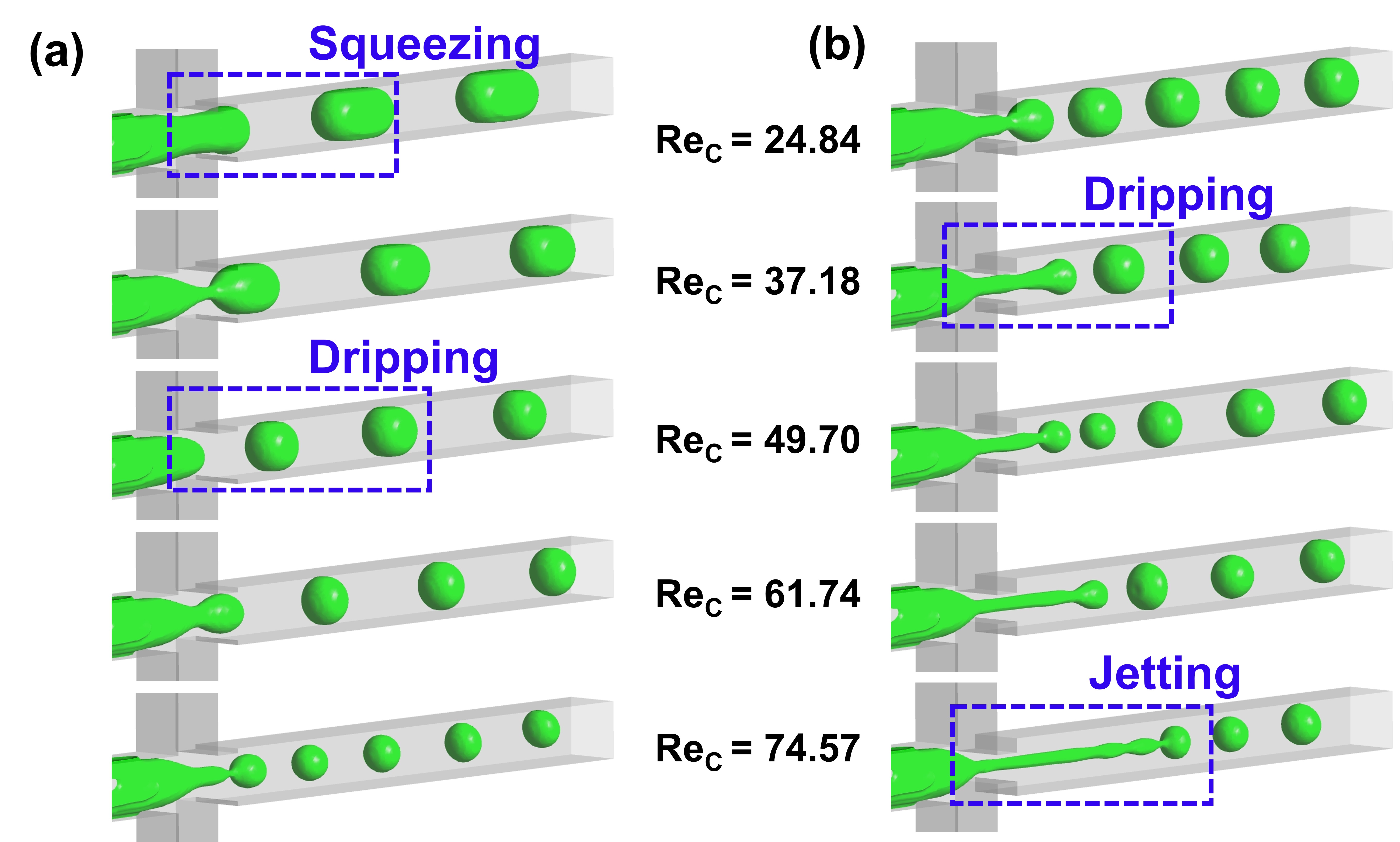}
	\caption{\label{fig:QCy11} Effect of continuous phase flow rate on droplet formation and flow regime transition (a) modified flow\textendash focusing microchannel with ${w^{*}_{or}}= 0.83$, and (b) modified flow\textendash focusing microchannel with ${w^{*}_{or}}= 0.5$  at a fixed operating condition of $\theta$= 120\textdegree, orifice length $l_{or}= 300~\mu m$,  oil viscosity $\mu_o$ = 0.53 mPa s, $\mu_o /\mu_w$ = 0.59, interfacial tension $\gamma$ = 5.37 mN/m, and $Q_o$= 400~$\mu$L/min. }
\end{figure*}
Two different microchannels are considered to understand the influence of continuous phase flow rate on the flow regimes at a fixed condition. In the case of  ${w^{*}_{or}}= 0.83$, with increasing the Reynolds number squeezing and dripping regimes are identified. Fig.~\ref{fig:QCy11}a shows that a transition between squeezing and dripping flow regimes is observed with increasing $Re_{c}$ from 24.84 to 49.70. Since the dispersed phase experiences more shear forces as $Re_{c}$ increases. Deformation of the interface mainly due to pressure differences or shear forces in the continuous phase causes a dispersed curve in the neck at the cross\textendash junction. It results in the pinch\textendash off of droplets from the dispersed phase flow. As a result of shear force, it leads to a significant change in droplet length and shape, as shown in Fig.~\ref{fig:QCy11}a. Similar flow transition from dripping to jetting  for lower  orifice width (i.e., ${w^{*}_{or}}= 0.5$ ) as shown in Fig.~\ref{fig:QCy11}b. With the increase, the $Re_{c}$ elongated jet is observed, and the pinch\textendash off position is shifted towards the middle of the channel. A Rayleigh \textendash Plateau instability in the dispersed thread may be responsible for propagating fluctuations in the elongated jet and strong, cohesive forces among the liquid molecules. \cite{rayleigh1879capillary,anjum2022microfluidic,wang2022perturbations} Interestingly the droplet shape changed from a smaller plug to near\textendash spherical for modified flow\textendash focusing microchannel with ${w^{*}_{or}}= 0.83$, specifically from the squeezing to the dripping regime. In the case of ${w^{*}_{or}}= 0.5$, near-spherical droplets are observed in both dripping and jetting regimes.

\begin{figure*}[]
	\centering
	\includegraphics[width=0.80\textwidth]{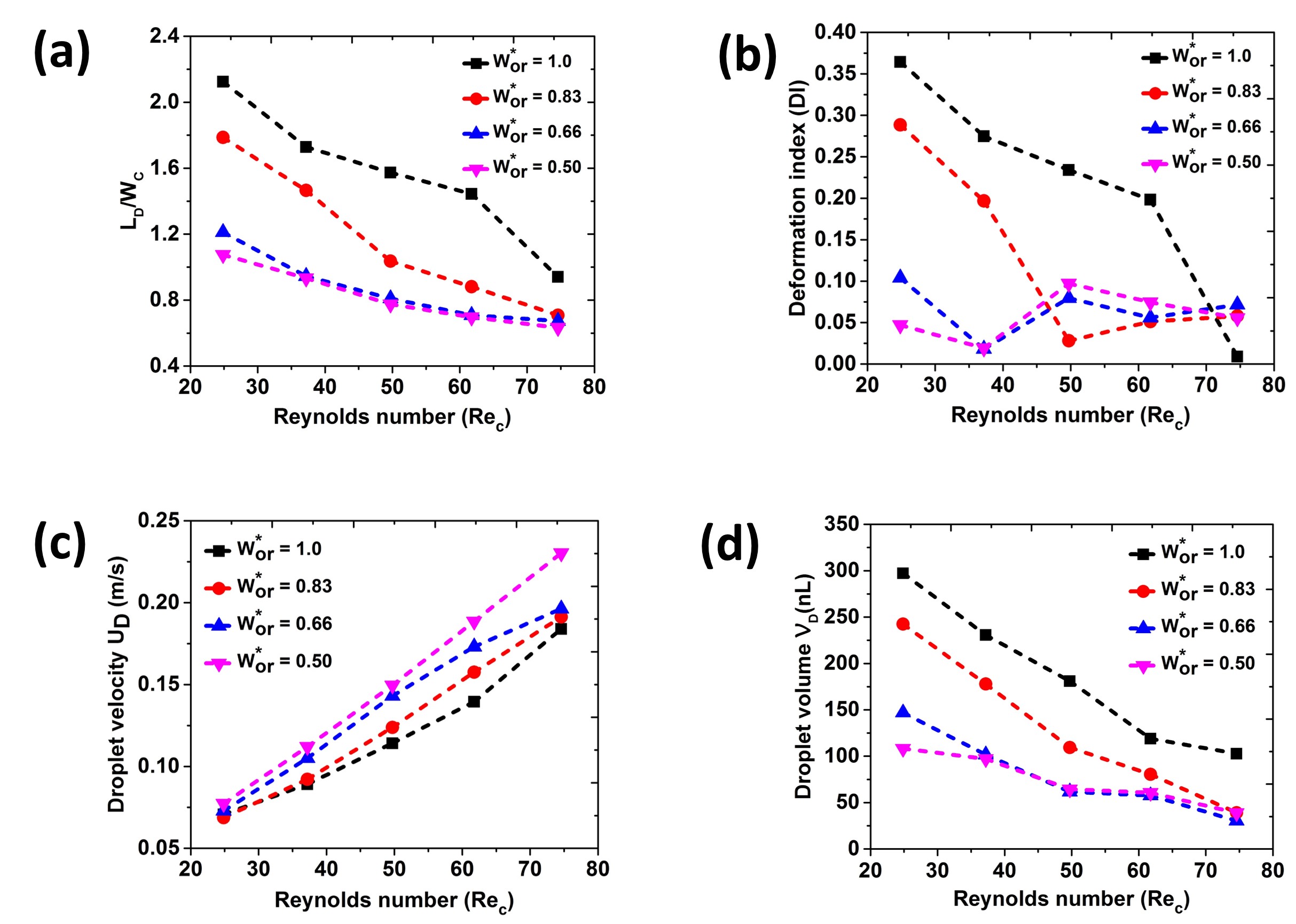}
	\caption{\label{fig:QC1} Effect of continuous phase flow rate on (a)  non\textendash dimensional droplet length, (b) droplet deformation index, (c) droplet velocity, and (d) droplet volume at a fixed operating condition of $\theta$= 120\textdegree, orifice length $l_{or}= 300~\mu m$,  oil viscosity $\mu_o$ = 0.53 mPa s, $\mu_o /\mu_w$ = 0.59, interfacial tension $\gamma$ = 5.37 mN/m, and $Q_o$= 400~$\mu$L/min. }
\end{figure*}
\begin{figure}[!h]
	\centering
	\includegraphics[width=0.50\textwidth]{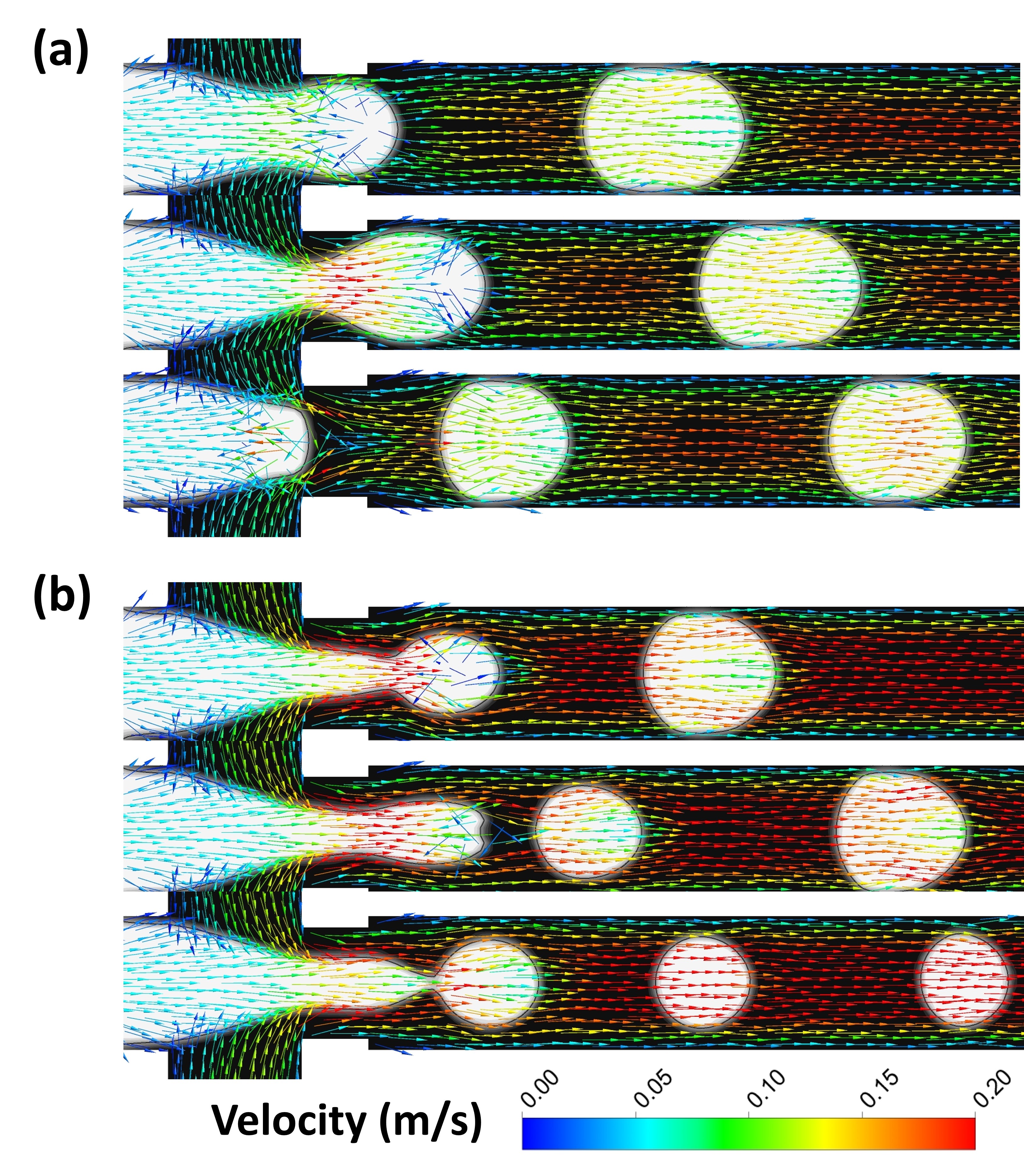}
	\caption{\label{fig:QCp11} Effect of continuous phase flow rate on velocity field distribution during droplet formation (a) $Re_{c}= 37.180$ and (b) $Re_{c}= 61.74$ at a fixed operating condition of $\theta$= 120\textdegree, ${w^{*}_{or}}= 0.83$ orifice length $l_{or}= 300~\mu m$,  oil viscosity $\mu_o$ = 0.53 mPa s, $\mu_o /\mu_w$ = 0.59, interfacial tension $\gamma$ = 5.37 mN/m, and $Q_o$= 400~$\mu$L/min. }
\end{figure}
In order to gain further insight into the droplet formation behavior is quantified based on $Re_{c}$. Non\textendash dimensional droplet length is found to decrease with an increase in $Re_{c}$ as shown in Fig.~\ref{fig:QC1}a. This is mainly attributed to the change in shear force on the dispersed phase. As indicated, within the normalized orifice width ${w^{*}_{or}}$ from 0.5 to 1.0, droplet length significantly changed by altering the $Re_{c}$. It is found that the inertial forces developed from the continuous phase play an important role in governing droplet formation in a modified flow\textendash focusing microchannel. Note that, when ${w^{*}_{or}=1}$ resulting standard flow\textendash focusing microchannel. Interestingly, droplet deformation index decreases with an increase in $Re_{c}$ for ${w^{*}_{or}} = 1$ as shown in Fig.~\ref{fig:QC1}b. However, with an increase in $Re_{c}$, the droplet height is smaller than its length for ${w^{*}_{or}} = 0.83, 0.66$, and $0.50$ due to flow transition from squeezing to dripping and dripping to jetting (see Fig.~\ref{fig:Ra12}a). In the jetting regime, the droplet height is higher than its length. Due to this fact, the DI for ${w^{*}_{or}} = 0.83, 0.66$, and $0.50$ showed different DI trends. Droplet velocities increased with an increase in $Re_{c}$. For smaller orifice width microchannel, droplet velocity is relatively higher than in other cases, as shown in Fig.~\ref{fig:QC1}c. This can be ascribed to changes in liquid film thickness and droplet shape. However, the droplet velocity change is negligible for lower  $Re_{c}$. In addition, droplet volume is also quantified (in Fig.~\ref{fig:QC1}d). The droplet volume is mainly decreases with an increase in $Re_{c}$ for all the cases.  

Fig.~\ref{fig:QCp11} demonstrates the velocity field distribution in the dripping regime. Velocity magnitude increases at the junction as the droplet enters the main channel due to blocking the dispersed phase at cross\textendash junction. Maximum velocity magnitude is observed around the neck of the dispersed phase thread and the middle of the microchannel. Recirculation patterns reveal that the droplet grows primarily in radial directions during the expansion phase, while the continuous phase obstructs its expansion. As shown in the Fig.~\ref{fig:QCp11}a, the inside velocity magnitude of the liquid plug is higher than the inside velocity magnitude of the droplet. Specifically, it is related to the shape of the droplet and the thickness of the film around it. The velocity re\textendash circulation inside the liquid plug is higher, but this observation is different for higher $Re_{c}$ even the flow regime is dripping as shown in Fig.~\ref{fig:QCp11}b. At the cross\textendash junction area, the velocity and the pressure inside the continuous phase increase, resulting in a change of neck curvature and a necking regime. As a result, the shear force strongly influences the droplet formation and pinch\textendash off position (Fig.~\ref{fig:QCp11}). 

\subsection{Effect of interfacial tension}
\begin{figure*}[]
	\centering
	\includegraphics[width=0.80\textwidth]{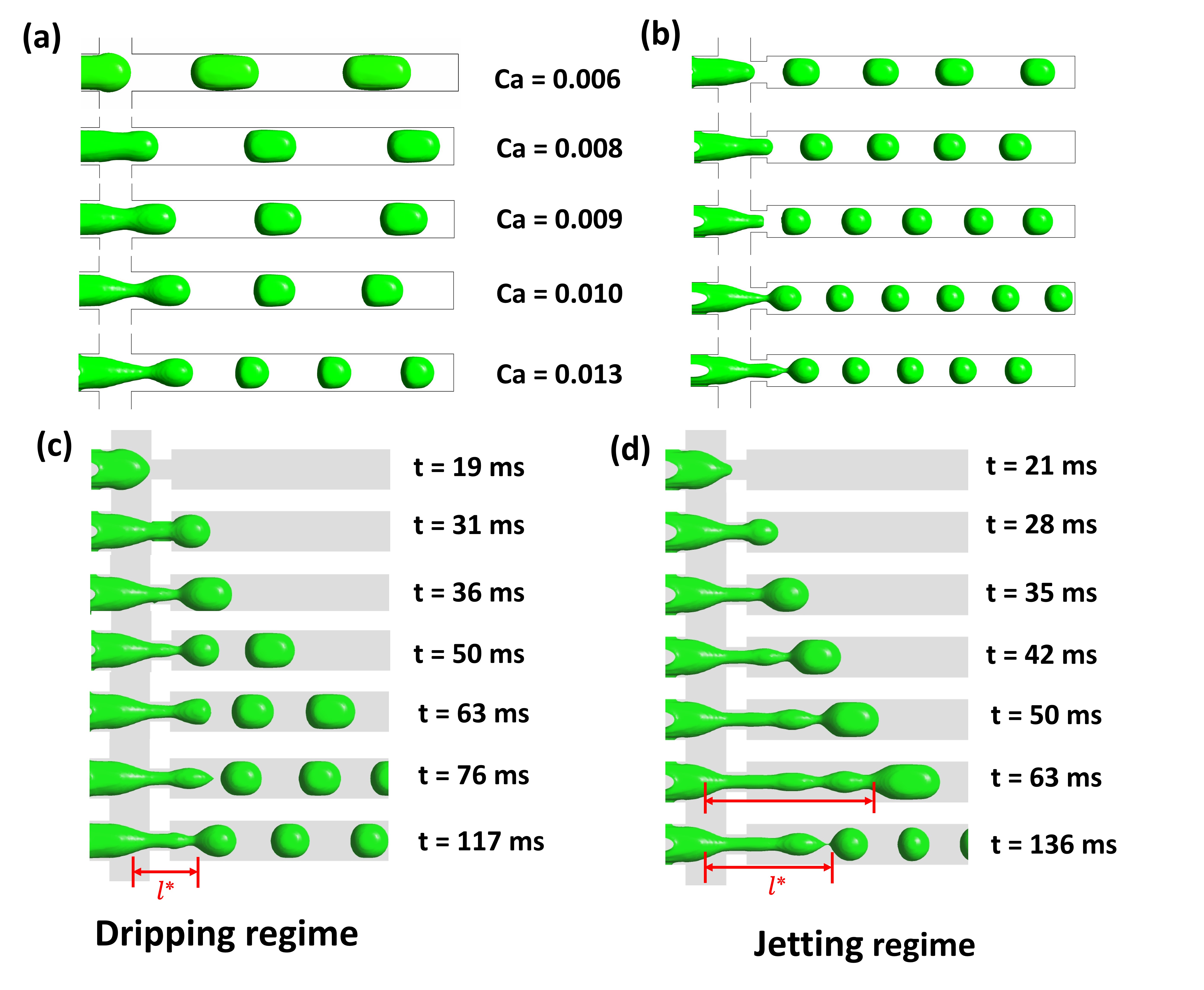}
	\caption{\label{fig:Ca1s} Effect of interfacial tension on droplet formation for (a) standard flow\textendash focusing microchannel without orifice ${w^{*}_{or}}=1$ and (b) modified flow\textendash focusing microchannel  ${w^{*}_{or}}= 0.5$. Flow regime transition for (c) $Ca=0.006$, ${w^{*}_{or}}= 0.5$ and (d) $Ca=0.013$, ${w^{*}_{or}}= 0.5$ at a fixed operating condition of $\theta$= 120\textdegree, orifice length $l_{or}= 300~\mu m$,  oil viscosity $\mu_o$ = 0.53 mPa s, $\mu_o /\mu_w$ = 0.59, interfacial tension $\gamma$ = 5.37 mN/m, $Q_w/Q_o=2$ and $Q_o$= 400~$\mu$L/min. }
\end{figure*}

In this section, the influence of interfacial tension has been systematically studied for different geometries at a fixed operating condition. The results are analyzed based on the capillary number $ \big(Ca = \frac{ U_{c}\mu_{c}}{\gamma}\big)$ by altering the interfacial tension. Droplet formation mechanism and flow regime have strongly been controlled by interfacial tension force and viscous forces.  From the experimental point of view, adding surfactants can alter interfacial tension.\cite{kovalchuk2018effect,fu-2012s,fu2009bubble} Typically, with increasing the surfactant concentration, the interfacial tension of the solution decreases. In the present study, interfacial tension is considered based on the practical range of an oil\textendash water system. \cite{yao2018formation} It can be seen from Fig.~\ref{fig:Ca1s}a,b that with increasing the $Ca$, droplet shape changes from plug to near\textendash spherical.
However, an increase in $Ca$ indicates the flow regime transition from squeezing to dripping for standard flow\textendash focusing microchannel ${w^{*}_{or}}=1$ and dripping to jetting for modified flow\textendash focusing microchannel ${w^{*}_{or}}=0.5$. Our finding on flow regime transition from dripping to jetting observation is good in agreement with the experimental work of \citet{lashkaripour2019performance}. As we can see in Fig.~\ref{fig:Ca1s}, the non\textendash dimensional droplet length is higher for lower $Ca$ and decreases with an in $Ca$. This is because interracial tension forces play a role in controlling the droplet size. Remarkably, a clear distinction in droplet length was observed for all the geometries. It is worth noting that when there are a low or moderate number of capillaries, drop formation is dominated by interfacial forces, which are mediated by the confinement of microchannels. Fig.~\ref{fig:Ca1s} c and d demonstrate the temporal evolution of droplet break\textendash in for different capillary numbers at a fixed orifice width. Droplet regimes shifted from dripping to jetting. The extended thread is observed in the jetting regime compared to the low $Ca$. Therefore,  the interfacial tension significantly impacts the flow regime's transition and droplet shape.

\begin{figure*}
	\centering
	\includegraphics[width=0.80\textwidth]{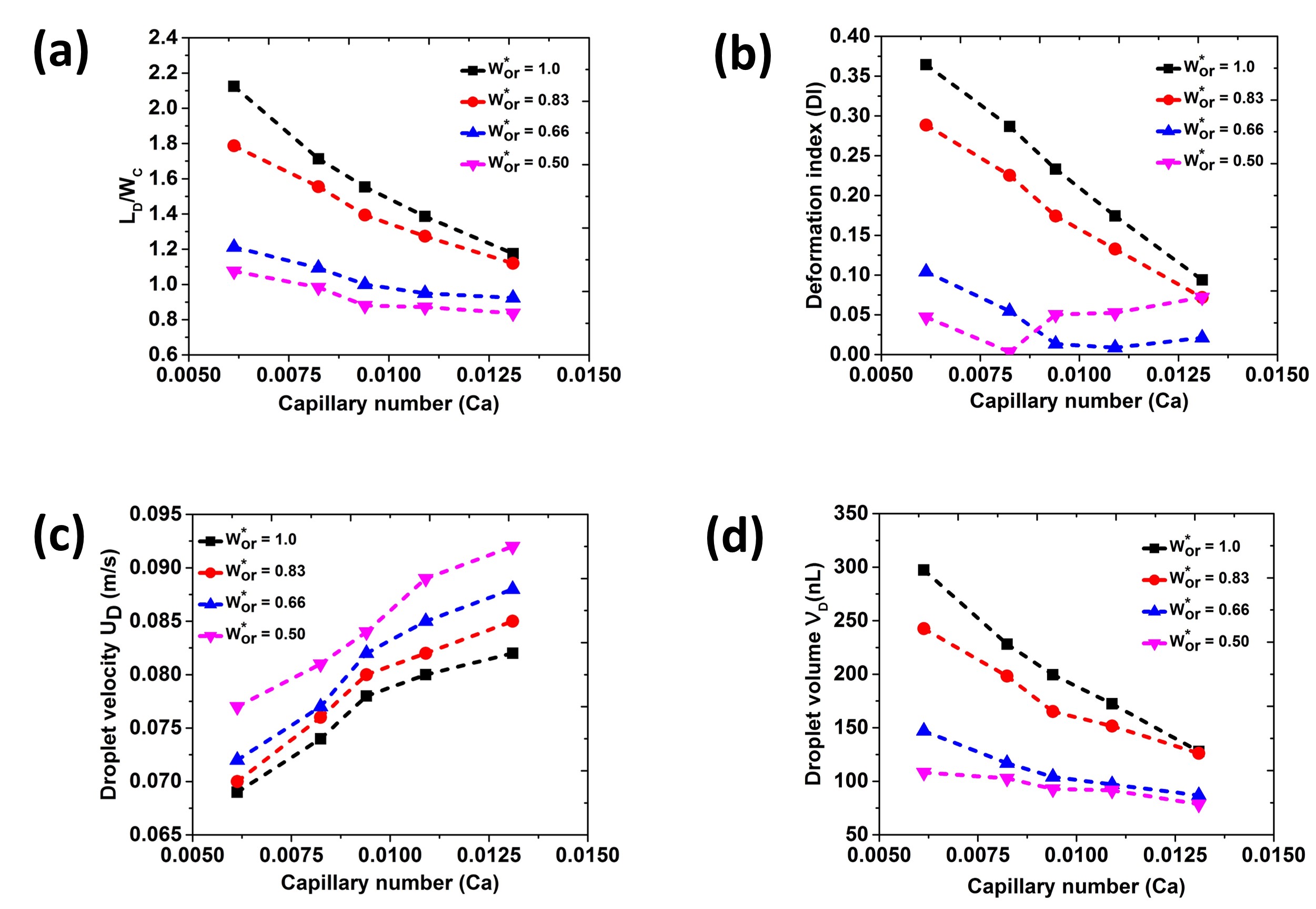}
	\caption{\label{fig:Ca1ss} Effect of interfacial tension on droplet formation (a)  non\textendash dimensional droplet length, (b) droplet deformation index, (c) droplet velocity, and (d) droplet volume at a fixed operating condition of $\theta$= 120\textdegree, orifice length $l_{or}= 300~\mu m$,  oil viscosity $\mu_o$ = 0.53 mPa s, $\mu_o /\mu_w$ = 0.59, interfacial tension $\gamma$ = 5.37 mN/m, $Q_w/Q_o=2$, and $Q_o$= 400~$\mu$L/min. }
\end{figure*}

Furthermore, a qualitative analysis was performed based on the capillary number (Ca) for different microchannels at a fixed orifice length. As shown in Fig.~\ref{fig:Ca1ss}a the non\textendash dimensional droplet length decreases significantly with increases in capillary number for all the microchannel configurations. 
In the case of a standard flow\textendash focusing microchannel, droplet length was relatively higher than in other cases. This phenomenon represents that interfacial tension significantly influences non\textendash dimensional droplet length. Similar to previous sections, a classification was considered to distinguish droplet shape based on deformation index (\textit{DI}). The results indicate that the droplet shape changes from an elongated plug to a small plug for the orifice width of  ${w^{*}_{or}}$= 1 and 0.5 microchannels. However, in the case of smaller orifice width, droplet shape changed from plug to near\textendash spherical, altering the $Ca$. Similar results were also reported on flow\textendash focusing microchannels in the literature.\cite{wu2017role}

 For higher $Ca$, droplets changed from a spherical shape. Interestingly,  ${w^{*}_{or}}= 0.66$ showed near\textendash spherical droplets with increase in $Ca$ as shown in Fig.~\ref{fig:Ca1ss}b. It is concluded that higher$Ca$ will result in a significant change in droplet shape and flow regime transition. The droplet velocity was also quantified for all the cases shown in Fig.~\ref{fig:Ca1ss}c. Droplet velocity increases with an increase in $Ca$. This is due to significant changes in liquid film thickness and droplet shape. It can be explained by the fact that the inertial force between the wall and the dispersed phase dominates when the liquid film thickness is higher from the droplet rear to the nose. In addition, droplet volume was also analyzed, as shown in Fig.~\ref{fig:Ca1ss}d. Interestingly, for $Ca$  higher, the droplet volume change is minimal for modified microchannels. This might be due to jetting regimes where small droplets are pinch\textendash off for the flow condition. 

\subsection{Flow regimes}
\begin{figure*}[]
	\centering
	\includegraphics[width=0.80\textwidth]{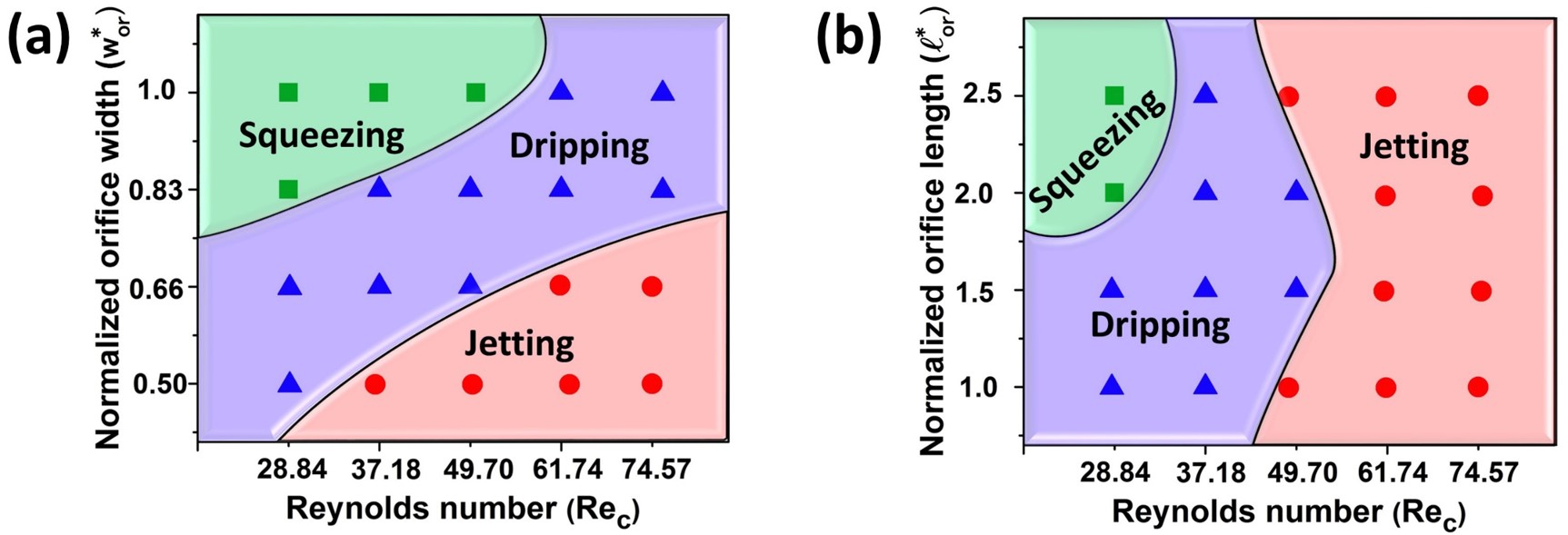}
	\caption{\label{fig:Ra1} Flow regime maps of droplet formation regime in different flow\textendash focusing microchannels as a function continuous phase Reynolds number with (a) normalized orifice width ${w^{*}_{or}}$ with orifice length of $l_{or}= 300~\mu m$ and (b)  normalized orifice length of ${l^{*}_{or}}$ with orifice width $w_{or}= 300~\mu m$ at a fixed operating condition of  $\theta$= 120\textdegree,  oil viscosity $\mu_o$ = 0.53 mPa s, $\mu_o /\mu_w$ = 0.59, interfacial tension $\gamma$ = 5.37 mN/m, and $Q_o$= 400~$\mu$L/min.}
\end{figure*}

We now further explored the role of geometric configurations on flow regime maps.  The numerically captured flow regimes are constructed in flow pattern maps. Fig.~\ref{fig:Ra1} shows the flow regime maps of an oil\textendash water system as a function of continuous phase Reynolds number and normalized orifice width/length. Three different flow regimes are identified for the range of operating conditions  and geometric configurations. Flow regimes are clearly distinguished by different colors, and flow transition is marked thick black color line. For the standard flow\textendash focusing microchannel  (i.e., ${w^{*}_{or}}$ = 1),  as shown in Fig.~\ref{fig:Ra1}a, with increasing the $Re_c$ number there are only two flow regimes found. For the low $Re_c$ squeezing regime is observed,  with increasing the $Re_c$, the transition regime appears. The squeezing regime forms more stable plug droplets, which are typically larger than microchannel  width (Fig.~\ref{fig:QCy11}a). However, with increasing $Re_c$ from $49.70$  to $74.57$ flow regime shifted from squeezing to dropping regime, and mono\textendash sized spherical droplets ($DI = 0$) are observed (Fig.~\ref{fig:QC1}b). It means that the droplet formation at higher flow range becomes more prominent to produce uniform droplets. Similarly, with decreasing the normalized orifice width, for low Reynolds number ($Re_c<49.70$) the flow transition is observed between the squeezing to the dripping regime. As is well known, the dripping and jetting regimes can produce spherical and small droplets than other regimes. However, the jetting regime produces less uniform droplets as its slender thread for the formation of satellite droplets. In this regime, droplets can form due to Rayleigh\textendash Plateau instability (i.e., capillary instability).\cite{rayleigh1879capillary,wang2022perturbations} 

From Fig.~\ref{fig:Ra1}a, it is found that for low $Re_c$, dripping regime is covered more area in flow regime map. By contrast, the jetting regime is a more dominant flow pattern at high $Re_c$, especially for smaller normalized orifice widths. Furthermore, another possibility of generalizing the flow regime maps based on normalized orifice length (${l^{*}_{or}}$) is also presented for different continuous phase $Re_c$ as shown in Fig.~\ref{fig:Ra1}b. It can be seen that  for low $Re_c$  squeezing and the dripping regimes are observed for different normalized orifice lengths. Due to a fixed orifice width, with increasing the $Re_c$ flow transition is observed from dripping/squeezing to the jetting regime as depicted in Fig.~\ref{fig:Ra1}b. In conclusion, the increase in normalized orifice length  for higher $Re_c$  can significantly impact on flow regime transition from dripping/squeezing to jetting. This is because of the increased length of the orifice, and the driving force exerted on the thread is not sufficient to pinch\textendash off the droplet at the cross junction. 
\begin{figure*}[]
	\centering
	\includegraphics[width=0.80\textwidth]{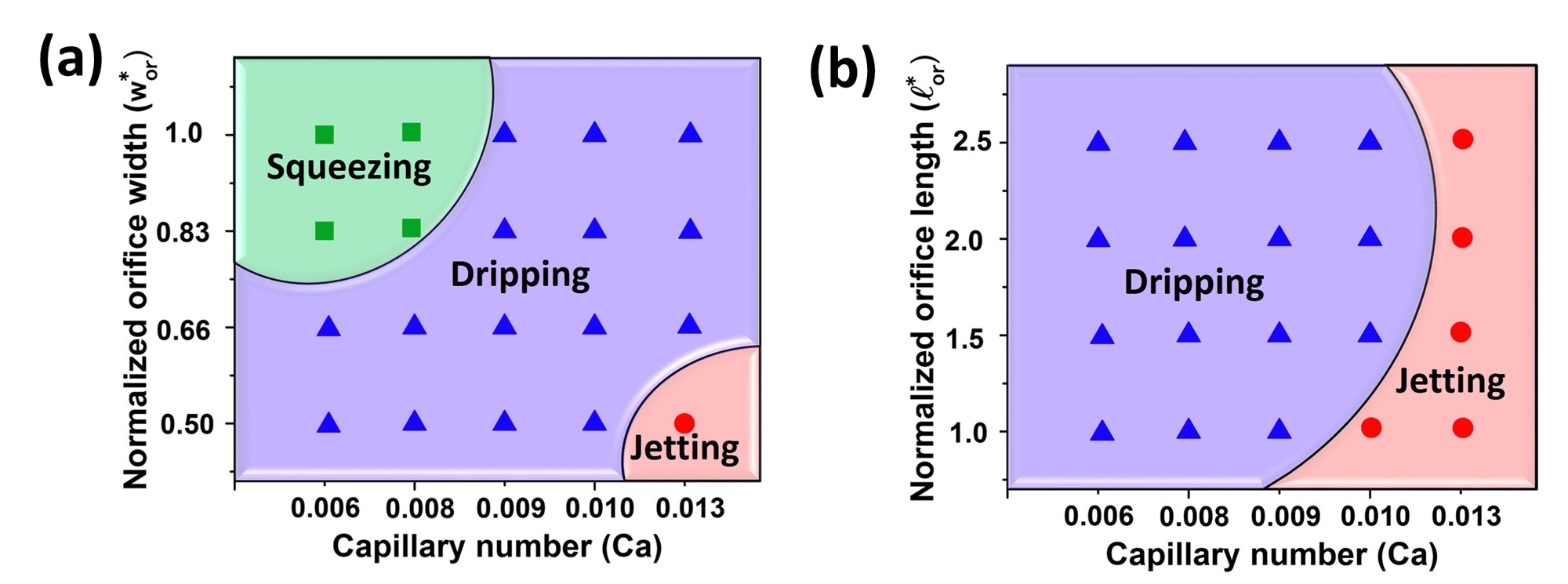}
	\caption{\label{fig:Ra12} Flow regime maps of droplet formation regime in different flow\textendash focusing microchannels as a function continuous phase capillary number with (a) normalized orifice width ${w^{*}_{or}}$ with orifice length of $l_{or}= 300~\mu m$, and (b)  normalized orifice length ${l^{*}_{or}}$ with orifice width of $w_{or}= 300~\mu m$ at a fixed operating condition of  $\theta$= 120\textdegree,  oil viscosity $\mu_o$ = 0.53 mPa s, $\mu_o /\mu_w$ = 0.59, $Q_w/Q_o=2$, and $Q_o$= 400~$\mu$L/min.  }
\end{figure*}
Furthermore, Fig.~\ref{fig:Ra12} shows the variation of $Ca$ from $0.006$ to $0.013$ on flow regimes for different normalized orifice width and length configurations. Fig.~\ref{fig:Ra12}a illustrates three different flow regimes that were observed for the considered range of $Ca$. Importantly, the dripping regime covers a large area in the flow maps. A large portion of the dripping regime indicates that the probability of obtaining more spherical droplets is higher in the range of operating condition considered. In the case of smaller $Ca$, decreasing the normalized orifice width plays a critical role in the flow transition from squeezing to dripping. As discussed above, droplet shape also changed the plug to spherical with an increase in $Ca$ for ${l^{*}_{or}}$ = 1\textendash 0.83 as shown in Fig.~\ref{fig:Ca1s}a. In the dripping regime, the dispersed phase is primarily influenced by viscous, pressure, interfacial, and inertial forces. Besides, for smaller normalized orifice width geometry configuration jetting regime is observed at higher $Ca>0.010$ (Fig.~\ref{fig:Ca1s}d). It can be noted that the variation of orifice width and interfacial forces lead to form squeezing and dripping regimes on the flow map, while the jetting regime is very limited. 
Fig.~\ref{fig:Ra12}b shows the flow regime map for different normalized orifice lengths. Two different flow regimes, such as dripping and jetting, are identified with altering the interfacial tension. For low capillary numbers, a dripping regime is observed for smaller normalized orifice length. With an increase in  Ca (i.e., $Ca<0.009$) flow regime remains stable for all the normalized orifice length configurations. However, the higher $Ca>0.010$ flow transition is observed from dripping to the jetting regime, and a similar observation also reported by \citet{lashkaripour2019performance}. At higher $Ca$, the low interfacial tension between oil and water during detachment of the droplet prevents expansion of the dispersed phase. For higher $Ca$, jetting is observed for all normalized orifice length configurations. Therefore, Fig.~\ref{fig:Ra12}b shows that geometry and capillary number should always be considered simultaneously when predicting droplet formation performance, especially in a dripping regime.  
\section{ SUMMARY AND CONCLUSIONS}
\label{conclusions}

In this work, we have numerically investigated the three\textendash dimensional two\textendash phase flow in a flow\textendash focusing microchannel with geometrical confinement. Geometrical confinement has been shown to play a significant role in droplet formation. A  3D unsteady CLSVOF method\textendash based CFD model was developed for droplet formation, and prediction is compared with literature data. This study demonstrated that the droplet formation and flow regime transition can be controlled by altering the geometrical confinement and operating condition. Effect of constriction width, length, continuous phase rate, and interfacial tension length on droplet size, velocity, shape, and volume systemically investigated. Non\textendash dimensional droplet length found increases with an increase in normalized orifice width. However, droplet velocity is decreased with an increase in normalized orifice width. As the normalized orifice length increased, droplet length decreased while droplet velocity increased. With increasing the normalized orifice width and length, squeezing and dripping flow regimes are observed at a fixed operating condition. Considerable refinement of the mesh near the solid wall can capture the liquid film's thinness.

For the first time, flow regime maps were also developed based on normalized orifice width, length, and non\textendash dimensional numbers to describe the flow regimes transition in a flow\textendash focusing microchannel. Three main pattern groups are observed: squeezing, dripping, and jetting. Notably, the flow transition from squeezing to dripping and dripping to jetting was identified by controlling the continuous phase flow rate and fluid interfacial tension. The flow regime maps provided insights to characterize the droplet formation regimes and mechanical for different geometrical confinement condition. The transitions in flow patterns were examined as a function of inertia, viscous shear, and interfacial tension. Besides, this work illustrates the droplet shape and formation frequency for various flow condition. 

The results of this study confirm that our CFD model can predict flow regimes in a modified flow\textendash microchannel. In summary, our findings provide a general idea of how geometrical confinement affects droplet formation. It is anticipated that the trends observed in this study will facilitate tuning for a high\textendash throughput production of mono\textendash sized droplets. Therefore, our findings are expected to provide a better understanding and practical guidelines for the applications, including drug delivery systems, emulsion formation, and various biotechnology applications.

  



\begin{acknowledgments}
	This work is supported by IIT Kharagpur under the Centre of Excellence for Training and Research in Microfluidics.
\end{acknowledgments}
\section*{Declaration of competing interest}
The authors declare that they have no known competing financial interests or personal relationships that could have appeared to influence the work reported in this paper.
\bibliography{POF}
\end{document}